\documentclass[twocolumn,preprintnumbers,superscriptaddress,amsmath,amssymb]{revtex4-2}
\usepackage{float}
\usepackage{graphicx}
\usepackage{dcolumn}
\usepackage{bm}
\usepackage{epstopdf}
\usepackage{xcolor}
\usepackage{textcomp}
\usepackage{soul}
\usepackage{csquotes}
\usepackage{amsbsy}
\usepackage{mathrsfs} 
\usepackage{amsmath}
\usepackage{bigints} 
\usepackage{amsthm,amssymb} 
\usepackage[utf8]{inputenc}
\usepackage[english]{babel}
\usepackage[shortlabels]{enumitem}
\usepackage{float}
\usepackage{mathrsfs,bigints,mathtools,dsfont}
\usepackage[colorlinks=true,linkcolor=blue,citecolor=blue]{hyperref}%
\usepackage[toc]{appendix}
\usepackage{longtable}

\newtheorem*{remark}{Remark}

\begin{document}
	
	\title{{Synchronization in repulsively coupled oscillators}}
	\author{Simin Mirzaei }\thanks{ Both are first authors of the article} \affiliation{Department of Biomedical Engineering, Amirkabir University of Technology (Tehran polytechnic), Iran }  
	\author{Md Sayeed Anwar}\thanks{ Both are first authors of the article} \affiliation{Physics and Applied Mathematics Unit, Indian Statistical Institute, 203 B. T. Road, Kolkata 700108, India} 
	\author{Fatemeh Parastesh}\affiliation{Department of Biomedical Engineering, Amirkabir University of Technology (Tehran polytechnic), Iran }
	\author{Sajad Jafari}\affiliation{Department of Biomedical Engineering, Amirkabir University of Technology (Tehran polytechnic), Iran } \affiliation{Health Technology Research Institute, Amirkabir University of Technology (Tehran polytechnic), Iran}
	\author{Dibakar Ghosh}\email{dibakar@isical.ac.in}\affiliation{Physics and Applied Mathematics Unit, Indian Statistical Institute, 203 B. T. Road, Kolkata 700108, India}

    \begin{abstract}
    	A long-standing expectation is that two repulsively coupled oscillators tend to oscillate in opposite directions. It has been difficult to achieve complete synchrony in coupled identical oscillators with purely repulsive coupling. Here, we introduce a general coupling condition based on the linear matrix of dynamical systems for the emergence of the complete synchronization in pure repulsively coupled oscillators. The proposed coupling profiles (coupling matrices) define a bidirectional cross-coupling link that plays the role of indicator for the onset of complete synchrony between identical oscillators. We illustrate the proposed coupling scheme on several paradigmatic two-coupled chaotic oscillators and validate its effectiveness through the linear stability analysis of the synchronous solution based on the master stability function approach. We further demonstrate that the proposed general condition for the selection of coupling profiles to achieve synchronization even works perfectly for a large ensemble of oscillators.                  		
    \end{abstract}
    
    \maketitle	
    
    \section{Introduction} 
     Dynamical networks have attracted much attention due to their exciting behaviors \cite{belykh2016bistability,dorfler2013synchronization,dorfler2014synchronization,olmi2015chimera}. One of the most important behaviors of dynamical networks is complete synchronization which occurs when all coupled oscillators have the same temporal evolution \cite{syn_book,boccaletti2002synchronization,arenas2008synchronization}. Synchronization in complex dynamical systems has become a hot topic in various fields \cite{pecora1990synchronization,cuomo1993circuit,heagy1994synchronous,rulkov1995generalized}. Synchronization plays an essential role in many natural phenomena or artificial applications \cite{perc2009optimal,anwar2022stability,mikhaylov2013sequential,belykh2004connection}. Heartbeat rhythm \cite{babaoglu2007firefly}, modern economic networks \cite{abraham2017chaotic}, neural networks \cite{milanovic1996synchronization}, and food web analysis \cite{blasius2000chaos} are some examples of the applications of synchronization in networks. Since the development of the master stability function (MSF) scheme \cite{pecora1998master}, which makes it possible to analyze synchronization in large oscillator networks with efficiency \cite{intra2}, the field of network synchronization has experienced explosive growth.
     \par It is important to note that the coupling between interacting oscillators has mostly been assumed to be positive (attractive) \cite{parastesh2019synchronizability,anwar2021enhancing,parastesh2022blinking}, which drives the oscillators to advance in the same direction and induces in-phase alignment. However, the repulsive (inhibitory) coupling is very common in biological systems such as  ensembles of inhibitory neurons \cite{hoppensteadt1997weakly}. Kim et al. \cite{kim2004pattern} studied a two-dimensional array of oscillators with phase-shifted coupling, which acts as a repulsive coupling in the network. In this regard, most of the studies on synchronization have considered the coupling between oscillators to be either solely attractive \cite{parastesh2019synchronizability,anwar2021enhancing,parastesh2022blinking,anwar2022stability,pecora1998master,pecora1990synchronization,huang2009generic,anwar2021relay} or mixed where both positive and negative coupling coexist simultaneously \cite{majhi2020perspective,rabinovich2006dynamical,saha2017coupling,saha2022resilience,dixit2019dynamics,dixit2020static,xu2021collective}. Yang et al. \cite{yang2019synchronization} investigated the influences of both positive and negative connections in the power networks and revealed some counterintuitive results. Leyva et al. \cite{leyva2006sparse} showed that synchronizing non-identical attractively coupled oscillators in a small-world network could be enhanced by considering a small fraction of phase-repulsive couplings.  K. Kovalenko et al. \cite{kovalenko2021contrarians} have investigated that synchronization can be achieved between repulsively coupled oscillators through the introduction of many-body interactions along with pairwise interactions. Nevertheless, the study of network synchronization with purely pairwise repulsive coupling, specifically the investigation of complete synchronization, has been overlooked as inhibitory coupling generally forces the oscillators to move apart and causes out-of-phase alignment \cite{kim2004pattern,elson1998synchronous}. Therefore, a natural question arises- when do coupled oscillators with purely pairwise repulsive coupling scheme achieve complete synchrony? To give a plausible answer to this question, here we introduce a general condition for the selection of an appropriate coupling scheme that yields complete synchrony in repulsively coupled oscillators. The proposed coupling condition defines a bidirectional cross-coupling link based on the off-diagonal elements of the linear matrix of the corresponding dynamical oscillators. The cross-coupling is defined here as the linear diffusive coupling involving two similar variables of the dynamical systems and added to the dynamics of a different state variable. An appropriate selection of this cross-coupling link yields complete synchrony in ensembles of dynamical oscillators with purely inhibitory coupling. The key characteristics of the coupling profile selection, particularly the insertion of specific cross-coupling connections, are demonstrated with the examples of two-coupled chaotic systems, namely the Hindmarsh-Rose (HR) neuron model, Chen system, R\"{o}ssler oscillators, Sprott system, and NE3 system, as dynamics of individual systems. The effectiveness of these coupling profile selections to predict the achievement of complete synchrony in repulsively coupled oscillators is validated using the linear stability analysis of synchronous solution based on the MSF formalism. The choice of appropriate coupling profile also works perfectly to predict the achievement of complete synchrony in large ensembles $(N>2)$ of repulsively coupled oscillators. 
     \par The remainder of this article is organized as follows. We first review the conventional MSF approach in Sec. \ref{Msf-review}, which plays an important role to guarantee the linear stability of the synchronous solution. Followed by this, in Sec. \ref{repulsive_examples}, we demonstrate the main results. We start with an example of two-coupled Hindmarsh-Rose neurons in Sec. \ref{hr_example}, to explain the mechanism of adding the appropriate coupling profile for the emergence of complete synchronization. This leads to the proposition of general condition for the selection of appropriate coupling profiles in many other coupled dynamical systems, detailed in Sec. \ref{general_condition}. We successfully extend the results to network motifs of three and four-nodes in Sec. \ref{motifs}. Furthermore, examples of larger networks, including $20$-node ring and random network of HR neurons, are illustrated. Finally, in Sec. \ref{conclusion}, we draw a conclusion by summarizing the results.               
 \section{Synchronization analysis based on MSF} \label{Msf-review}
	 The master stability function (MSF) is a common approach for finding the necessary conditions for stable synchronization in a dynamical network consisting of identical coupled oscillators. The MSF calculation is entirely independent of the network topology. This method is briefly reviewed in the following.
	 We consider a network consisting of $N$ coupled identical oscillators with $\mathbf{x}_{i}$ being the $m$-dimensional vector of the $i$th oscillator. The dynamical behavior of each oscillator can be described by $\dot{\mathbf{x}}_{i}=F(\mathbf{x}_{i})$, where $F(\mathbf{x}_{i})$ is the velocity field. The dynamical equation of the network of $N$ coupled oscillators can be defined as,
	 \begin{equation}\label{msf_eq1}
	 	\begin{array}{l}
	 	 \dot{\mathbf{x}}_{i}=F(\mathbf{x}_{i})-\epsilon\sum\limits_{j=1}^{N} G_{ij} h(\mathbf{x}_{j}),	
	 	\end{array}
	 \end{equation}    
	 where $\epsilon$ is the coupling strength, $G$ is the Laplacian of the network connectivity matrix, and $h(\mathbf{x}_{j})$ is the coupling function between oscillators. The coupling function $h(\mathbf{x}_{j})$ can be written as $h(\mathbf{x}_{j})= H\mathbf{x}_{j}$ for linear coupling functions, where $H$ is the coupling matrix that shows which state variables are involved in the coupling. 
	 The synchronization manifold of the network can be expressed as $\mathbf{x}_{1}=\mathbf{x}_{2}=\cdots=\mathbf{x}_{N}=\mathbf{s}$, where the synchronized solution $\mathbf{s}(t)$ satisfies $\dot{\mathbf{s}}=F(\mathbf{s})$. To evaluate the stability of the synchronization manifold, a perturbation can be considered as $\mathbf{y}_{i}(t)=\mathbf{x}_{i}(t)-\mathbf{s}(t)$. Tending the perturbations to zero guarantees the stability of the synchronization manifold $(\mathbf{s})$, and then, all of the oscillators approach the synchronization manifold. The dynamical equation of the perturbations can be written as,
	 \begin{equation}\label{msf_eq2}
	 	\begin{array}{l}
	 		\dot{\mathbf{y}}_{i}=[DF(\mathbf{s})-\epsilon\sum\limits_{j=1}^{N} G_{ij} Dh(\mathbf{s})] \mathbf{y}_{i},
	 	\end{array}
	 \end{equation}
	 where $DF(\mathbf{s})$ and $Dh(\mathbf{s})$ are the Jacobian matrices of functions $F$ and $h$ evaluated at $\mathbf{s}(t)$.
	 Using the eigenvalues of the matrix $G$, the variational equation of the network can be converted to decoupled systems. By applying the transform $\eta=Q^{-1}\mathbf{y}$, where matrix $Q$ is constructed from the eigenvectors of the matrix $G$, the decoupled form of Eq. \eqref{msf_eq2} can be obtained as
	 \begin{equation}\label{msf_eq3}
	 	\begin{array}{l}
	 		\dot{\eta}_{i}=[DF(\mathbf{s})-\epsilon\lambda_{i} Dh(\mathbf{s})] \eta_{i},
	 	\end{array}
	 \end{equation}     
	 where $\eta_{i}$ defines the variations of the $i$th oscillator, and $\lambda_{i}$ $(i=1,2,\cdots,N)$ are the eigenvalues of matrix $G$. The first eigenvalue for a connected network is zero (i.e., $\lambda_{1}=0)$ and the corresponding variational equation is along the synchronization manifold. Other eigenvalues $\lambda_{i}>0$, $i=2,3,\cdots,N$ determine the stability of the variational equation and the synchronization manifold. Letting $K$ be the normalized coupling parameter defined as $K=\sigma \lambda$, the largest Lyapunov exponent of Eq. \eqref{msf_eq3} $(\Lambda(K))$ is the MSF. The negative MSF demonstrates that the synchronous manifold is stable.	 
	 \section{Results}\label{repulsive_examples}
	 In this section, we first consider a particular example on two-coupled Hindmarsh-Rose neuronal models and show that complete synchronization emerges using repulsive coupling. We verify our result using master stability function and basin stability analysis. Then we propose a general coupling scheme for other systems based on linear matrix of the isolated dynamical systems for complete synchronization using purely repulsive coupling and finally we extend the study for network motifs.  
    \subsection{Coupled Hindmarsh-Rose neurons} \label{hr_example}
         \begin{figure}[ht] 
    	\centerline{
    		\includegraphics[scale=0.45]{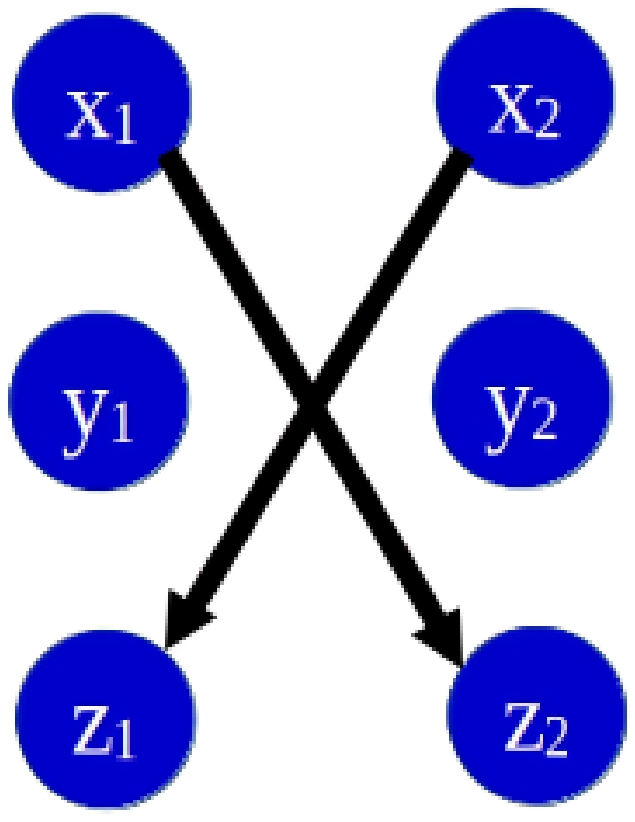}
    		\includegraphics[scale=0.32]{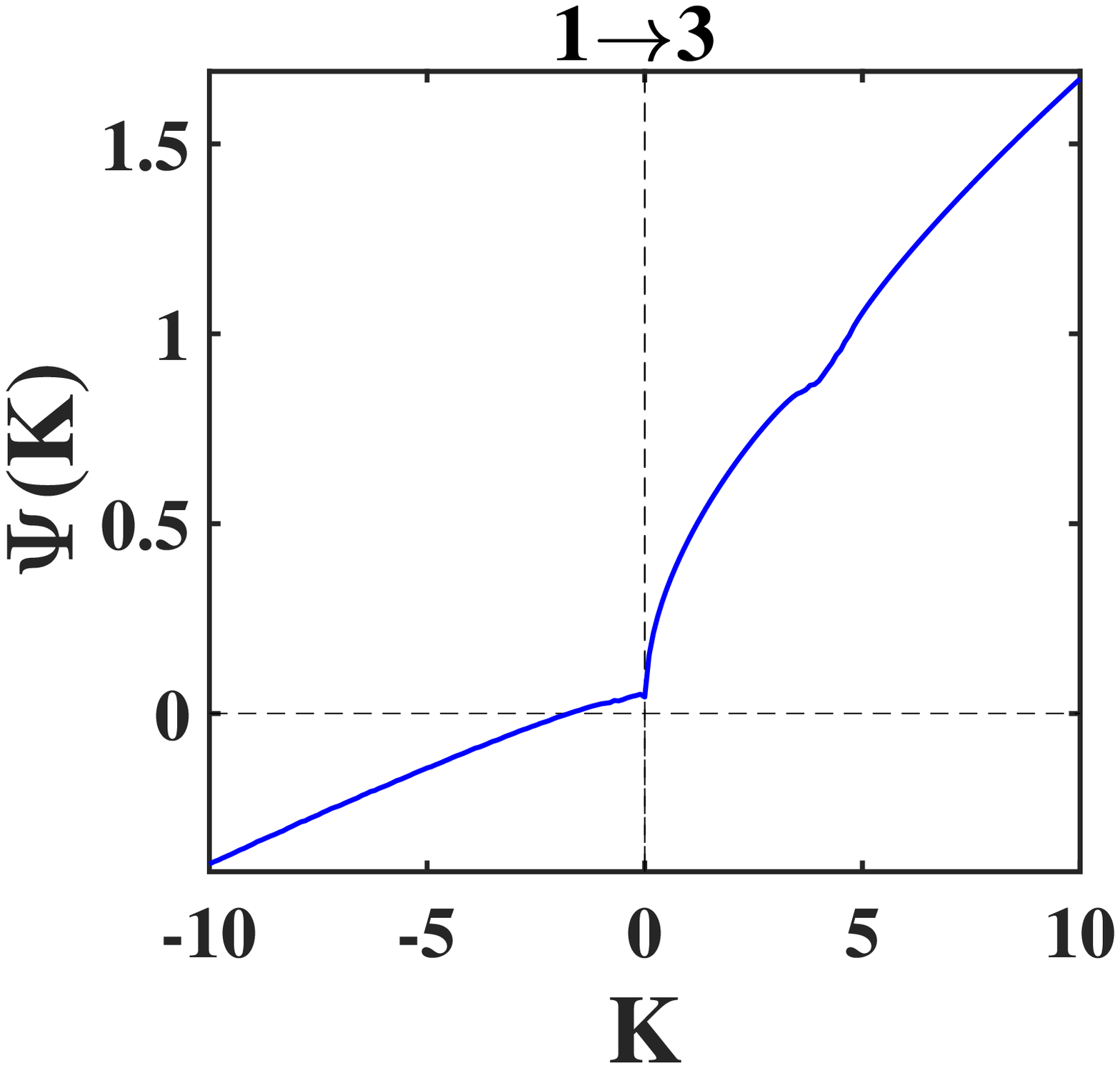}}
    	\caption{ (Left panel) Schematic representation of bidirectional $1\rightarrow 3$ coupling profile. The coupling function, including the $x_{1,2}$ variable, is added to the dynamics of the $z_{2,1}$. (Right panel) MSF of two-coupled Hindmarsh-Rose neuronal system in Eq. \eqref{hr_neuron} versus the normalized coupling parameter $K=2\epsilon$ for $1 \to 3$ coupling configuration that shows the occurrence of complete synchronous solution in the negative coupling regime.}
    	\label{hr_msf}
    \end{figure} 
    We start by considering two identical HR-neurons coupled with each other through a bidirectional cross-coupling configuration $1 \rightarrow 3$, i.e., the coupling is on the $1$st variable and added to the dynamics of the $3$rd variable (The coupling configuration is demonstrated schematically in the left panel of Fig. \ref{hr_msf}). Then the equation of motion governing the dynamics of the coupled system is given by,   
    \begin{equation}\label{hr_neuron}
    	\begin{array}{l}
    		\dot{x}_{1,2}=y_{1,2}+3 {x}_{1,2} ^ {2} - {x}_{1,2} ^ {3} -z_{1,2}+I , \\
    		\dot{y}_{1,2}=1-5 {x}_{1,2} ^ {2} -y_{1,2}, \\
    		\dot{z}_{1,2}= -rz_{1,2}+rs(x_{1,2}+1 . 6) + \epsilon (x_{2,1}-x_{1,2}),
    	\end{array}
    \end{equation}
    where the subscripts $(1,2)$ indicate the oscillators, and $I=3.2$, $r=0.006$, $s=4$ are system parameters that yield chaotic behavior of uncoupled HR-neurons. $\epsilon$ $(>,\; \mbox{or} < 0)$ is the real-valued constant that represents the strength of attraction or repulsion between the coupled neurons. Since our main objective is to find the region of synchronization when the neurons are repulsively coupled, we vary the coupling strength $\epsilon$ from negative to positive regime and investigate the locally stable synchronization state based on MSF approach. Figure \ref{hr_msf} (right panel) shows the MSF of coupled HR-neuron system in Eq. \eqref{hr_neuron} versus the normalized coupling parameter $K(=2\epsilon)$. One can notice that the synchronization is achievable for $K < - 1.8$, but no synchrony emerges in the positive regime of the coupling strength. Therefore, with the considered cross-coupling configuration $1\rightarrow 3$, two HR neurons can achieve a stable synchronized state when they repel each other with sufficient strength. 
    \begin{figure}[ht] 
    	\centerline{
    		\includegraphics[scale=0.35]{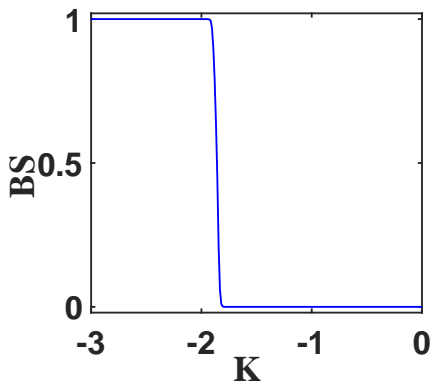}}
    	\caption{Basin stability (BS) of two-coupled Hindmarsh-Rose neuronal system in Eq. \eqref{hr_neuron} as a function of the normalized coupling parameter $K=2\epsilon$ for $1 \to 3$ coupling configuration. Here $BS=0$ indicates unstable synchronization state and $BS=1$ represents globally stable synchronization state.}
    	\label{hr_basin}
    \end{figure}
    \par Further, we carried out the basin stability measure \cite{menck2013basin,rakshit2017basin}, which quantifies the volume of the basin leading to the synchronized state, to see if the attained synchronous solution in the negative coupling regime is only stable for modest perturbation from the synchronization manifold. So, we choose $\mathcal{Q}_{0}=10^{4}$ different initial states distributed randomly over the phase space volume of uncoupled HR neuron, to quantify the fraction of states leading to the synchronized state. If $\mathcal{Q}$ number of states finally arrive at the synchronization state, then the basin stability of the synchronous state is measured as $BS=\frac{\mathcal{Q}}{\mathcal{Q}_{0}}$. The range of BS is $[0,1]$, where $BS = 0$ denotes that the synchronous state is unstable under any initial states, and $BS = 1$ denotes that it is globally stable under any nonlocal perturbation. If $0<BS<1$, the value of BS is the likelihood of restoring the synchronized state from perturbation for a typical initial state. Figure \ref{hr_basin} depicts the BS for the complete synchronous state, under the variation of normalized coupling strength $K(=2\epsilon)$. As observed, for $K < -1.91$, the value of BS is unity, which reflects that the complete synchronization solution is globally stable. Therefore, with the prescribed cross-coupling configuration, two HR neurons can achieve a globally stable synchronous state in the negative coupling regime.      
    \par Now, this specific choice of coupling profile for the HR-neuronal system can be made in a structured manner to achieve stable synchronization under repulsive coupling from the constant matrix that represents the linear part of the system, and consequently, we can construct generic coupling conditions for the appropriate choice of coupling profiles to accomplish complete synchrony in many pure repulsively coupled dynamical systems. In this context, the evolution of any dynamical system can be represented as, 
      \begin{equation}\label{general_eqn}
      	\begin{array}{l}
      		\dot{\mathbf{x}}= F(\mathbf{x})=L\mathbf{x}+g(\mathbf{x})+P,
      	\end{array}
      \end{equation}
  where ${\mathbf{x}} \in \mathbb{R}^{d}$, ($d=3$ for our considered systems) is the state vector, $L$ is the $d \times d$ constant matrix attributing to the linear part of the system, $g: \mathbb{R}^{d} \to \mathbb{R}^{d}$ accounts for the nonlinear part of the system, and $P$ is a $d \times 1$ constant matrix. For the HR-neuronal system,        
      \begin{equation}
      	\begin{array}{l}
      	  	L=\begin{pmatrix} 
      			0 & 1 & -1 \\
      			0 & -1 & 0 \\
      			rs & 0 & -r    	
      		\end{pmatrix}, g(X)=\begin{pmatrix} 
      		3x^2-x^3  \\
      		-5x^2 \\
      		0     	
      	\end{pmatrix}, P=\begin{pmatrix} 
      	I  \\
      	1 \\
      	1.6rs     	
      \end{pmatrix}.
      	\end{array}
      \end{equation}
    From the linear matrix $L$ of the HR-neuron model, we can observe that two nonzero elements- one positive and negative element $(L_{12} \; \mbox{and} \; L_{13})$ exist in the upper triangle of the linear matrix $L$ which are connected to the dynamics of 1st variable $\dot{x}_{1,2}$, and one positive element $(L_{31})$ exists in the lower triangle connected to the dynamics of 3rd variable $\dot{z}_{1,2}$. Moreover, the off-diagonal conjugate elements $L_{13}$ and $L_{31}$ are of opposite sign. This suggests the inclusion of bidirectional cross-coupling between two neurons defined by a coupling function involving $x_{1,2}$ variables to the dynamics of $z_{1,2}$ as described in the Eq. \eqref{hr_neuron} since the positive element is connected to $x_{1,2}$ variable in $\dot{z}_{1,2}$ and the negative element is connected to $z_{1,2}$ variable in $\dot{x}_{1,2}$.  
    \subsection{General coupling condition for two-coupled systems} \label{general_condition}
    The above observation with the HR-neuronal model allows us to introduce a general condition for the selection of an appropriate coupling profile between two repulsively coupled dynamical systems from their corresponding linear matrix $L$. For a three-dimensional system $(d=3)$, there are nine linear coupling configurations which can be defined as $1\to 1$, $1\to 2$, $1\to 3$, $2\to 1$, $2\to 2$, $2\to 3$, $3\to 1$, $3\to 2$ and $3\to 3$, where the notation $i \to j$ describes that the coupling is on the $i$th state variables and added to the $j$th state variables. Now the question is which coupling configuration is appropriate to achieve a complete synchronization state when the systems are repulsively coupled? To answer this question, we introduce a generic procedure that the nonzero off-diagonal conjugate elements determine the choice of a suitable coupling profile. Our observation shows that mostly the bidirectional cross-coupling between two negatively coupled identical dynamical systems is responsible for the emergence of complete synchrony. Therefore, we suggest the following coupling criteria:
    	 \begin{itemize}
    	 	\item { Suppose a nonzero element appears in the upper triangle of the linear matrix $L$, i.e., $L_{ij} \ne 0 $ $(i<j)$, where $i,j=1,2,3$ and its corresponding conjugate element $L_{ji}$ in the lower triangle is also non-zero and of opposite sign. Then depending on the sign of $L_{ij}$ and $L_{ji}$, a bidirectional cross-coupling is introduced between the systems. If the element in the upper triangle is negative and its conjugate element in the lower triangle is positive (i.e., $L_{ij}<0$ $(i<j)$ and $L_{ji}>0$), then a bidirectional cross-coupling link $i \rightarrow j $ is essential. Opposite coupling configuration ($j\rightarrow i)$ is made when $L_{ij}>0$ $(i<j)$ and $L_{ji}<0$.}
    	 \end{itemize}
    \par The statement above adequately validates our choice of bidirectional cross-coupling between two-coupled HR neurons, as shown in the previous section. We provide four more examples in support of our statement with paradigmatic Chen system \cite{chen1999yet}, Sprott-I system \cite{sprott1994some,*sprottI}, R\"{o}ssler oscillators \cite{rossler1976equation,*rossler} and  NE3 systems \cite{jafari2013elementary,*ne3}.  Further examples are demonstrated in Appendix \ref{appendix}. The parameters for all these dynamical systems are taken in such a manner that yields chaotic behavior for the uncoupled systems. The linear matrices for the aforementioned four systems are now arranged below, from left to right, respectively,     
    \begin{equation}\label{linear_flow_matrices}
    	\begin{array}{l}
    		L=\begin{pmatrix} 
    			-\alpha & \alpha & 0 \\
    			c-\alpha & c & 0 \\
    			0 & 0 & -\beta    	
    		\end{pmatrix}; \begin{pmatrix} 
    			0 & -0.2 & 0 \\
    			1 & 0 & 1 \\
    			1 & 0 & -1    	
    		\end{pmatrix}; \\\\ ~~~~~~~~~~
    		\begin{pmatrix} 
    			0 & -1 & -1 \\
    			1 & a & 0 \\
    			0 & 0 & -c    	
    		\end{pmatrix};
    		\begin{pmatrix} 
    			0 & 1 & 0 \\
    			0 & 0 & 1 \\
    			0 & -1 & 0    	
    		\end{pmatrix}.      
    	\end{array}
    \end{equation}  
     For the Chen system, the element $L_{12}$ in the upper triangle of the linear matrix $L$ is positive in sign, and its conjugate element $L_{21}$ in the lower triangle is negative for the choice of system parameter. Hence a bidirectional cross-coupling link involving $y_{1,2}$ is to be added to the dynamics of first variables $x_{1,2}$. From the linear matrix $L$ of the Sprott-I system, we can observe that the element $L_{12}$ is negative in sign and $L_{21}$ is of positive sign. Therefore, according to the proposition, a bidirectional cross-coupling involving $x_{1,2}$ variables is to be added to the dynamics of $y_{1,2}$ for the achievement of synchrony. For the R\"{o}ssler oscillator, the $L_{12}$ element of the linear matrix $L$ is $-1$ and the corresponding cross diagonal element $L_{21}$ is $1$, which are of opposite sign. So, a bidirectional cross-coupling link containing $x_{1,2}$ is to be appended to the evolution of the variables $y_{1,2}$. In a similar manner, from the linear matrix $L$ of the NE3 system, one can find that $L_{23}=1$ and $L_{32}=-1$, which are clearly of opposite signs. Therefore, a bidirectional link including $z_{1,2}$, added to the dynamics of 2nd variables $y_{1,2}$, yields a complete synchrony state in two repulsively coupled NE3 systems.
     \par As a result, we can obtain coupling matrix $H$ to construct repulsively coupled Chen system, Sprott-I system, R\"{o}ssler oscillator, and NE3 system that can achieve complete synchronization. The coupling matrices for all these four systems are represented below, from left to right respectively,           
     \begin{equation} \label{coupling_matrices}
     	\begin{array}{l}
     		H=\begin{pmatrix} 
     			0 & 1 & 0 \\
     			0 & 0 & 0 \\
     			0 & 0 & 0    	
     		\end{pmatrix};
     		
     		\begin{pmatrix} 
     			0 & 0 & 0 \\
     			1 & 0 & 0 \\
     			0 & 0 & 0    	
     		\end{pmatrix}; \\\\ ~~~~~~~~
     		
     		\begin{pmatrix} 
     			0 & 0 & 0 \\
     			1 & 0 & 0 \\
     			0 & 0 & 0    	
     		\end{pmatrix};
     		
     		\begin{pmatrix} 
     			0 & 0 & 0 \\
     			0 & 0 & 1 \\
     			0 & 0 & 0    	
     		\end{pmatrix}.
     	\end{array}
     \end{equation}   
The elements of these coupling matrices in Eq. \eqref{coupling_matrices} are selected correspondingly:
\begin{enumerate}
	\item Chen system: The coupling matrix contains only one non-zero element, $H_{12}=1$. This corresponds to the bidirectional cross-coupling configuration $2\rightarrow 1$. 
	\item Sprott-I system: The coupling matrix contains only one non-zero element, $H_{21}=1$, which accounts for the bidirectional cross-coupling configuration $1\rightarrow 2$.
	\item R\"{o}ssler oscillator: The coupling matrix contains only one non-zero element, $H_{21}=1$. This corresponds to the bidirectional cross-coupling configuration $1\rightarrow 2$.
	\item NE3 system: The coupling matrix contains only one non-zero element, $H_{23}=1$, attributing to the bidirectional cross-coupling configuration $3\rightarrow 2$.	 
\end{enumerate}
\begin{remark}
At this stage, it is important to note that we have proposed a general condition for the choice of one possible coupling profile that yields complete synchrony between coupled oscillators with pure repulsive coupling. There may be other suitable coupling schemes for the emergence of complete synchronization in repulsively coupled dynamical systems. Furthermore, our proposition is not able to predict the circumstances under which perfect synchrony is not achievable with pure repulsive coupling.  	
\end{remark}
 \begin{figure}[ht] 
	\centerline{
		\includegraphics[scale=0.22]{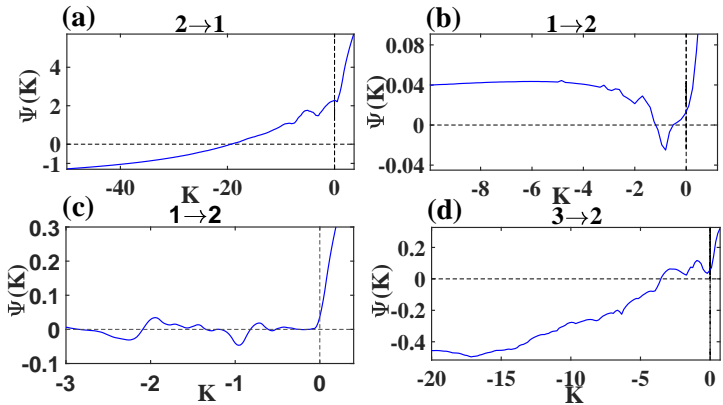}}
	\caption{ Variation of MSF as a function of normalized coupling strength $K=2\epsilon$ for four different two-coupled systems. (a) Chen system: coupling configuration $2 \rightarrow 1$; parameter values $\alpha=35$, $\beta=\frac{8}{3}$, $c=28$. (b) Sprott-I system: coupling configuration $1 \rightarrow 2$. (c) R\"{o}ssler oscillator: coupling configuration $1 \rightarrow 2$; parameter values $a=0.2$, $b=0.2$, $c=5.7$. (d) NE3 system: coupling configuration $3\rightarrow 2$.  }
	\label{4_system_lyap}
\end{figure}
 Now, in order to validate these coupling profiles for the emergence of complete synchrony, we perform the linear stability of the synchronization state based on the MSF scheme. Figure \ref{4_system_lyap} depicts the MSF of two-coupled Chen system, Sprott-I system, R\"{o}ssler oscillator and NE3 system respectively, by varying the normalized coupling strength $K(=2\epsilon)$ from negative to positive regime. In Fig. \ref{4_system_lyap}(a), for Chen system, the synchronization is achievable for $K<-19.25$ with $2 \rightarrow 1$ coupling configuration, which clearly validates our prediction based on the linear matrices about the emergence of synchronization state in negative coupling regime. Similarly, for all the other three systems, synchrony is achievable in the negative coupling region for our predicted coupling profiles based on the elements of linear matrices $L$. This is certainly validated by their respective MSF plots represented in Fig. \ref{4_system_lyap}(b)-\ref{4_system_lyap}(d). Furthermore, Fig. \ref{4_system_lyap} suggests that the MSFs for these systems with considered coupling configurations exhibit different types of functional behavior. For instance, the MSF $\Psi(K)$ for Chen system and NE3 system is a monotone decreasing function that intercepts the abscissa once at some critical coupling $K_{c}<0$, resulting in an unbounded region of synchronization. Additionally, for the Sprott-I system and R\"{o}ssler oscillator $\Psi(K)$ is not monotonic but intercepts the abscissa for two or more instances that yield bounded regions of synchronization. In case of the Sprott-I system, the MSF admits negative value in some range $(K_2,K_1)$, where $K_2<K_1<0$ and for R\"{o}ssler oscillator $\Psi(K)$ possesses more than two finite crossing points across the abscissa. Therefore, analogous to the previous studies with purely attracting coupling scheme \cite{huang2009generic}, for an appropriate choice of repulsive coupling configurations, two-coupled dynamical systems exhibit an unbounded and bounded region of stable synchronization state.           
   \subsection{Network Motifs} \label{motifs}  
  Now, we concentrate on network motifs, the fundamental units of many real-world networks, to see if the proposed condition for the choice of coupling configurations is effective in promoting complete synchrony in the negative coupling regime. In a network of $N$ oscillators, the dynamics of a node $i$ under bidirectional coupling configuration can be given as
  \begin{equation} \label{general_network_eq}
  	\begin{array}{l}
  	   \dot{\mathbf{x}}_{i}=F(\mathbf{x}_{i})+ \epsilon \sum\limits_{j=1}^{N}\mathscr{C}_{ij} H(\mathbf{x}_{j}-\mathbf{x}_{i}), \; i=1,2,\cdots,N.	
  	\end{array}
  \end{equation}         
  Here, $F(\mathbf{x}_{i})$ represents the isolated node dynamics with $\mathbf{x}_{i}$ being the $d=3$-dimensional state vector, $\epsilon$ is the strength of bidirectional coupling between any two nodes. The $N\times N$ symmetric matrix $\mathscr{C}$ denotes the connection topology of the network; $\mathscr{C}_{ij}=1$ if any two nodes $i$ and $j$ are connected through a bidirectional link and zero otherwise. As usual, $H_{(d \times d)}$ corresponds to the coupling matrix that defines through which variables a pair of nodes are connected with each other. Complete synchronization in network \eqref{general_network_eq} occurs when each node advances with the rest nodes in unison. The corresponding  synchronization manifold is defined by $\mathcal{S}=\{\mathbf{x}_{1}(t)=\mathbf{x}_{2}(t)=\cdots=\mathbf{x}_{N}(t)=s(t)\}$ and the dynamics of the synchronization solution $(x(t),y(t),z(t))=s(t)$ is determined by the associated uncoupled oscillator.       
  \par By using the Linear matrix $L$ of each node, we appropriately choose the coupling matrix $H$ between any two nodes in the network.   
  We now demonstrate the broad applicability of our proposed coupling condition through a series of examples on few network motifs that achieve complete synchrony in negative coupling regime. 
  \subsubsection{Three-node Chen systems}
   \begin{figure}[ht] 
  	\centerline{
  		\includegraphics[scale=0.12]{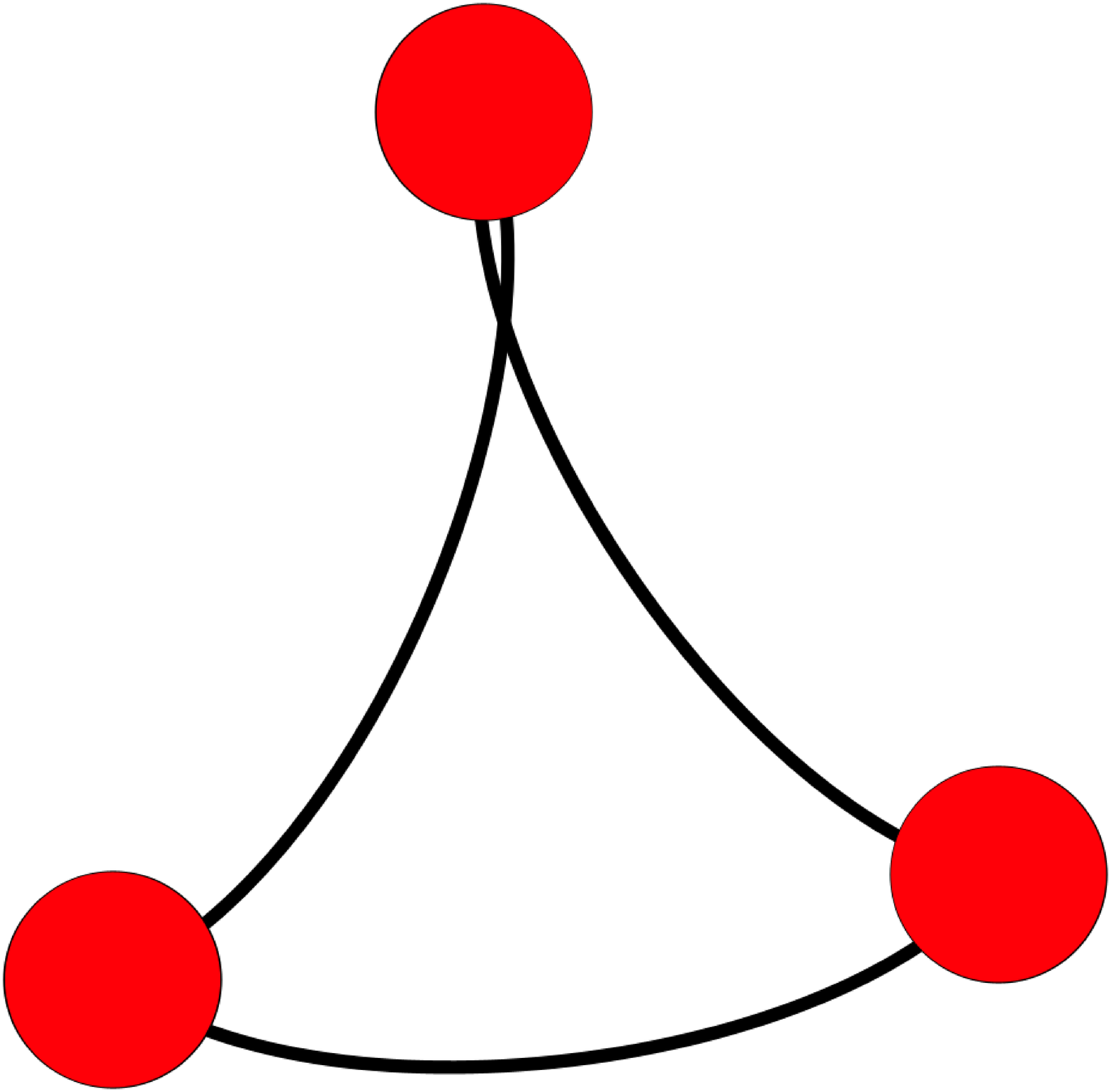}
  		\includegraphics[scale=0.27]{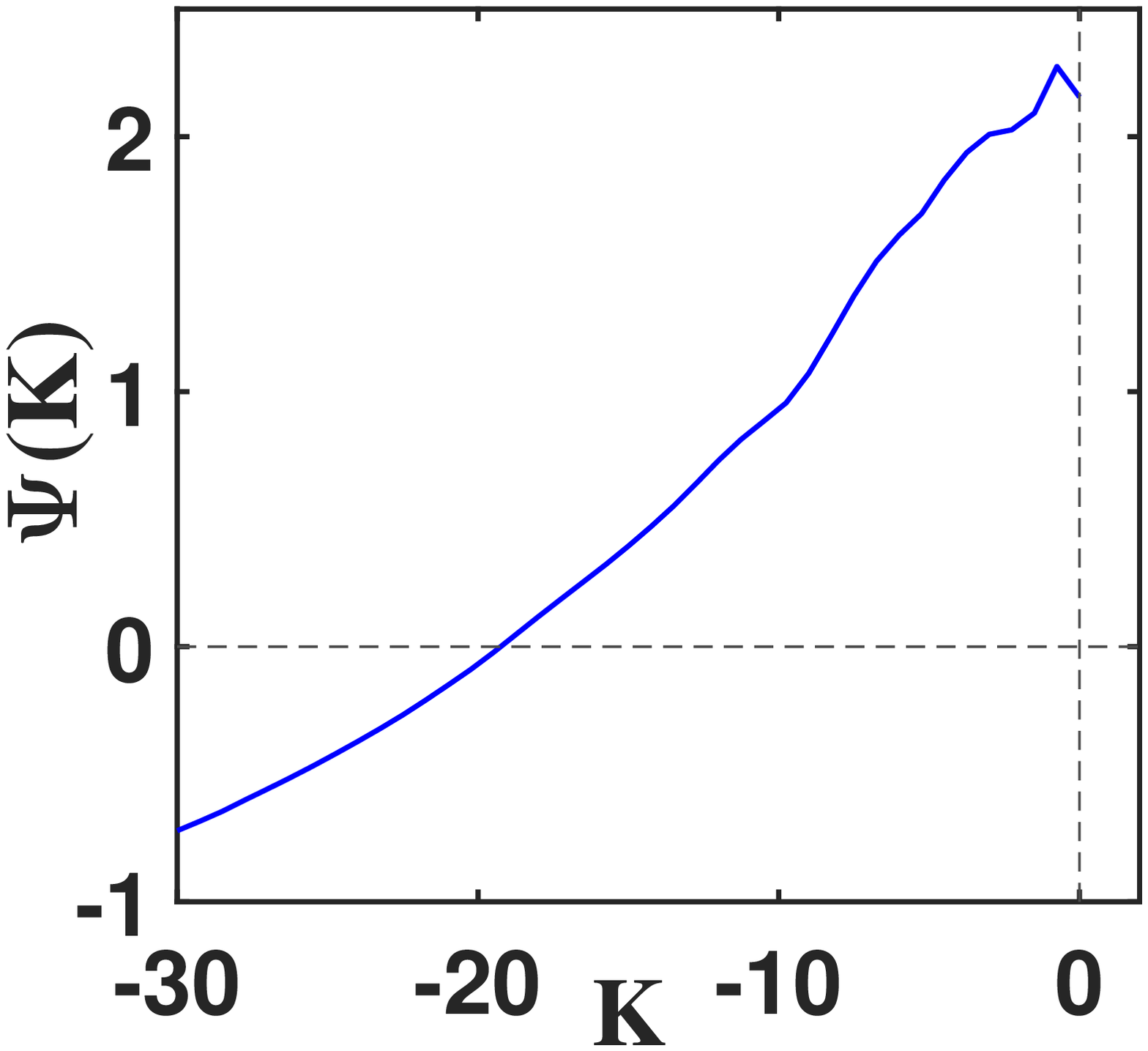}}
  	\caption{ (Left column) Schematic representation of a $N=3$ nodes global network. (Right column) MSF of coupled Chen system in Eq. \eqref{chen_network} as a function of the normalized coupling parameter $K=N\epsilon$ for $2 \to 1$ coupling configuration.}
  	\label{chen_k3_lyap}
  \end{figure}
  As a starter, we consider a global network of three nodes whose individual node dynamics is given by chaotic Chen system. The network connectivity matrix $\mathscr{C}$ and the corresponding coupling matrix $H$ are,
  \begin{equation}
  	\begin{array}{l}
  		\mathscr{C}=\begin{pmatrix}
  			0 & 1 & 1 \\
  			1 & 0 & 1 \\
  			1 & 1 & 0
  		\end{pmatrix}; \;H=\begin{pmatrix}
  		0 & 1 & 0 \\
  		0 & 0 & 0 \\
  		0 & 0 & 0
  	\end{pmatrix}.
  	\end{array}
  \end{equation}
  The equation of motion governing the dynamics of node $i$ can then be written as follows,
  \begin{equation}\label{chen_network}
  	\begin{array}{l}
  		\dot{x}_{i}=\alpha(y_{i}-x_{i}) + \epsilon \sum \limits_{j=1}^{N} \mathscr{C}_{ij}(y_{j}-y_{i}), \\
  		\dot{y}_{i}=(c-\alpha-z_{i})x_{i}+cy_{i}, \\
  		\dot{z}_{i}= x_{i}y_{i}-\beta z_{i},
  	\end{array}
  \end{equation}
  where $i=1,2,3$ and $\alpha=35$, $c=28$, $\beta=\frac{8}{3}$ are standard parameters that yield chaotic behavior in isolated Chen system. Now to validate the effectiveness of our proposed coupling profile based on the linear matrix $L$ of the uncoupled system, we perform the local stability of the synchronized solution based on the MSF approach.    
 Figure \ref{chen_k3_lyap} displays the MSF as a function of normalized coupling strength $K=3\epsilon$. The acquired MSF $\Psi(K)$ is a monotone decreasing function that crosses the abscissa once after a critical coupling strength $K=-19.25$. Beyond $K<-19.25$, the MSF remains always negative and results in an unbounded synchronization region.
 \subsubsection{Four-node HR neuron models}
  \begin{figure}[ht] 
 	\centerline{
 		\includegraphics[scale=0.12]{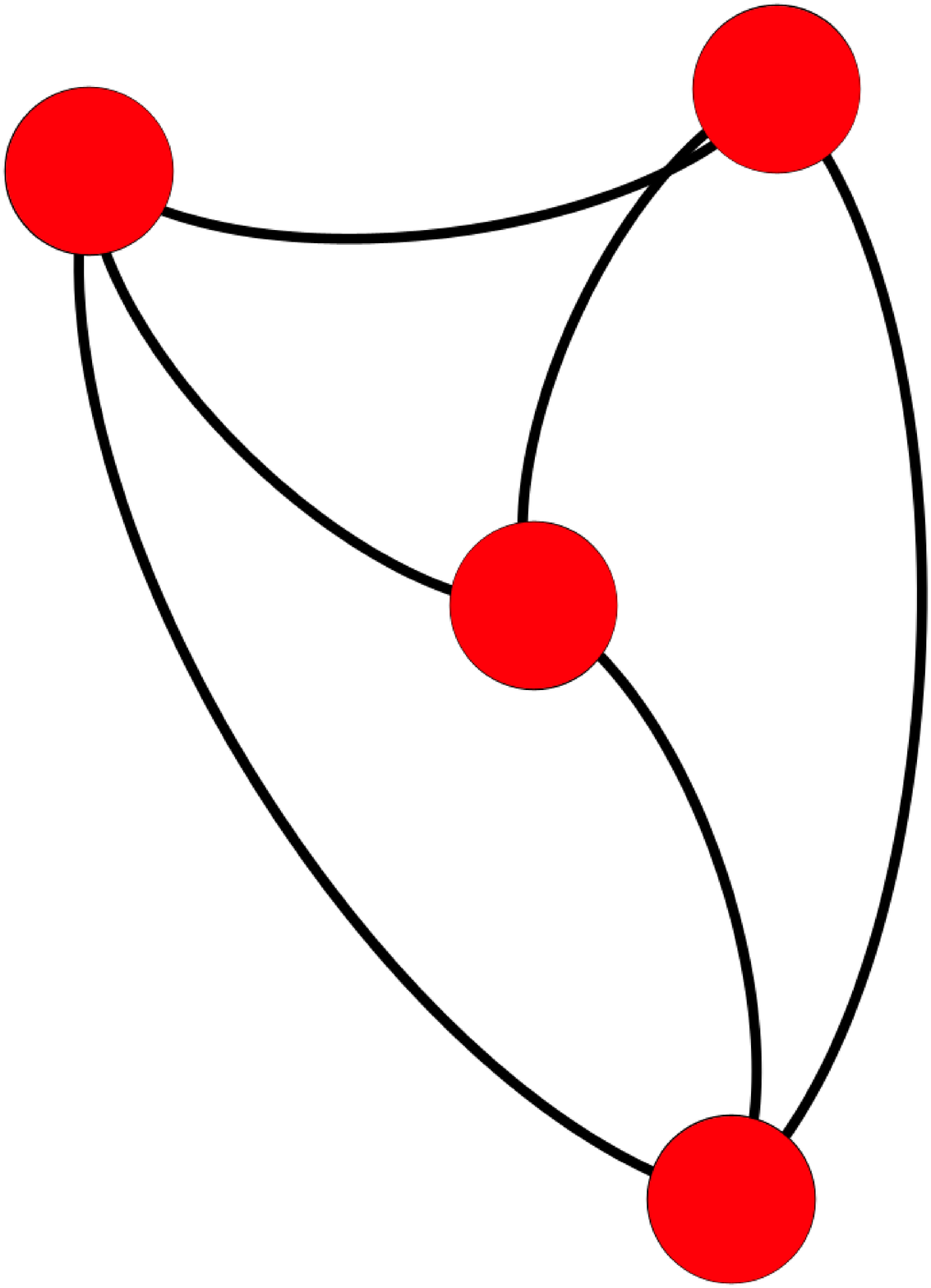}
 		\includegraphics[scale=0.27]{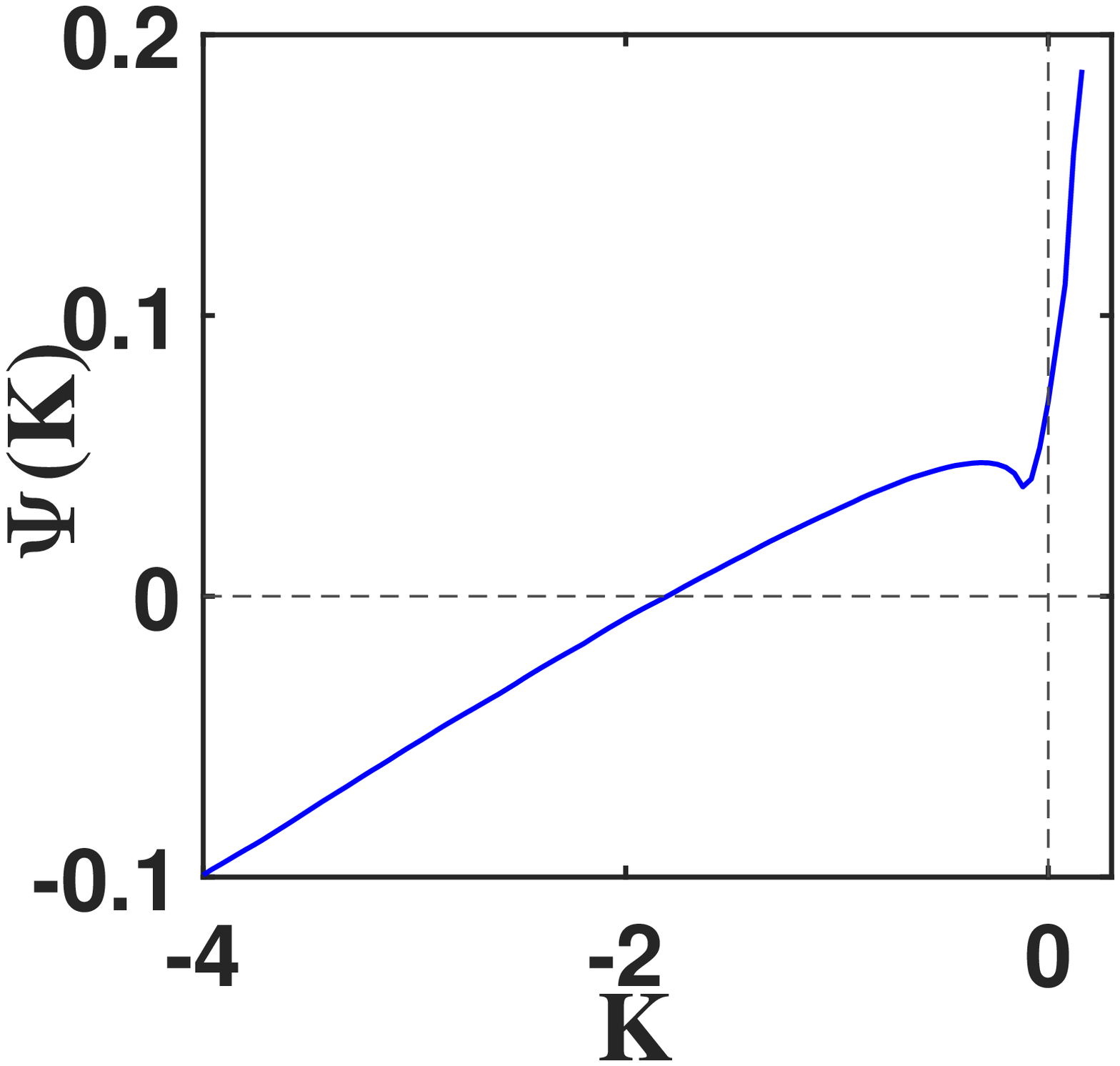}}
 	\caption{ (Left column) Pictorial representation of a global network with $N=4$ nodes. (Right column) MSF of coupled Hindmarsh-Rose neuronal system in Eq. \eqref{hr_network} versus the normalized coupling parameter $K=N\epsilon$ for $1 \to 3$ coupling configuration.}
 	\label{hr_k4_lyap}
 \end{figure}
  A globally coupled network of four HR neurons is considered as a second example. The network connectivity matrix $\mathscr{C}$ and the coupling matrix $H$ obtained from the proposed criteria based on the linear matrix $L$ of isolated HR-neuronal model are given by,      
    \begin{equation}
    	\begin{array}{l}
    		\mathscr{C}=\begin{pmatrix}
    			0 & 1 & 1 & 1 \\
    			1 & 0 & 1 & 1\\
    			1 & 1 & 0 & 1 \\
    			1 & 1 & 1 & 0
    		\end{pmatrix}; \;H=\begin{pmatrix}
    			0 & 0 & 0 \\
    			0 & 0 & 0 \\
    			1 & 0 & 0
    		\end{pmatrix}.
    	\end{array}
    \end{equation}
   The equation of motion governing the dynamics of node $i$ can then be written as follows,
 \begin{equation}\label{hr_network}
 	\begin{array}{l}
 		\dot{x}_{i}=y_{i}+3x^{2}_{i}-x^{3}_{i}-z_{i}+I, \\
 		\dot{y}_{i}=1-5x^{2}_{i}-y_{i}, \\
 		\dot{z}_{i}= -rz_{i}+rs(x_{i}+1.6)+ \epsilon \sum\limits_{j=1}^{N} \mathscr{C}_{ij} (x_{j}-x_{i}),
 	\end{array}
 \end{equation}
 where $i=1,2,3$ and $I=3.2$, $r=0.006$, $s=4$ are standard parameters that yield chaotic behavior in isolated HR-neuron model. Now to validate the efficacy of our proposed coupling profile based on the linear matrix of the uncoupled system, we execute the local stability of the synchronous solution based on the MSF approach.    
 Figure \ref{hr_k4_lyap} displays the MSF as a function of normalized coupling strength $K=4\epsilon$. The acquired MSF $\Psi(K)$ is a monotone decreasing function that crosses the abscissa once after a critical coupling strength $K=-1.85$. Beyond $K<-1.85$, the MSF remains always negative and results in an unbounded synchronization region.  
 \subsubsection{Ring of HR neurons}
 \begin{figure}[ht] 
 	\centerline{
 		\includegraphics[scale=0.12]{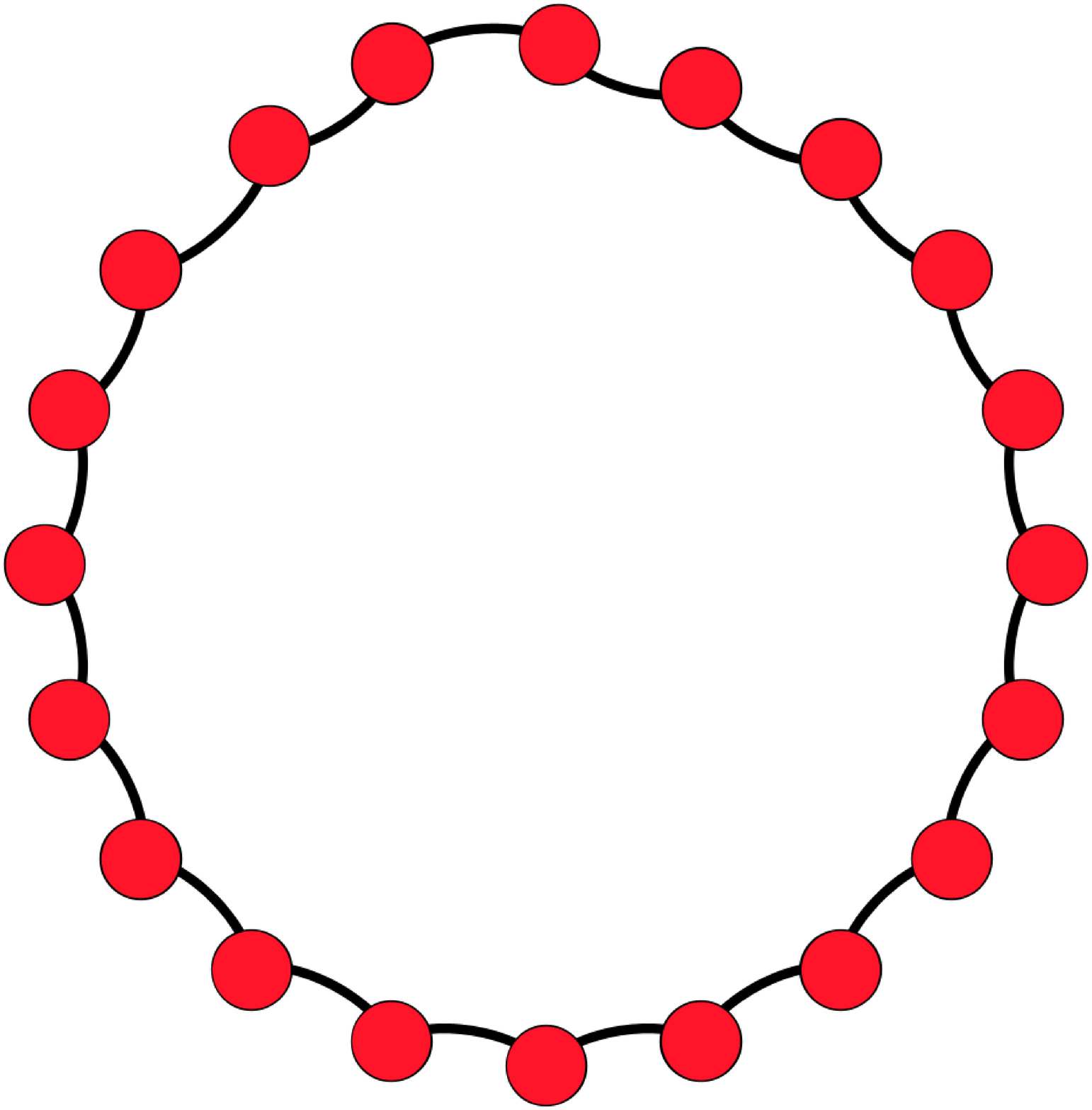}
 		\includegraphics[scale=0.27]{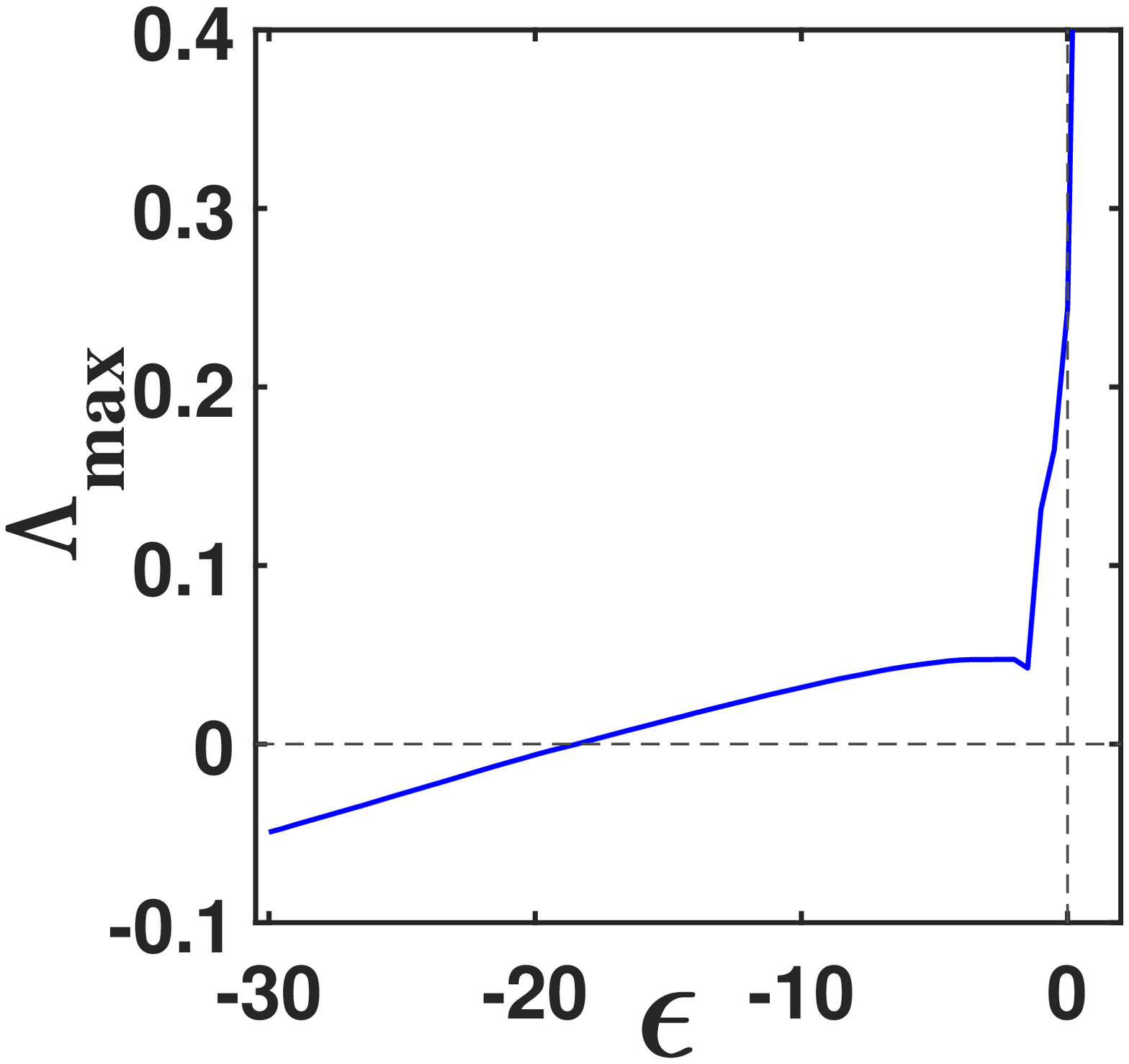}}
 	\caption{ (Left panel) Schematic representation of a ring of $N=20$ nodes. (Right panel) Variation of maximum transverse Lyapunov exponent $\Lambda_{max}$ as a function of coupling strength $\epsilon$ in the negative regime with $1 \rightarrow 3$ coupling configuration is displayed in solid blue line. For each individual HR neuron, the system parameters are $r=0.006$, $s=4$, and $I=3.2$.}
 	\label{hr_ring_lyap}
 \end{figure}
 Next, we consider a ring of $N=20$ nodes where the dynamics of each node are governed by chaotic HR neuronal model (a pictorial representation of network connectivity is given in the left panel of Fig. \ref{hr_ring_lyap}). Following our proposition based on the linear matrix $L$ of HR model, in this scenario also we consider the coupling scheme between any pair of nodes to be $1 \rightarrow 3$, i.e., a bidirectional cross-coupling link including the $x$-variables is added to the dynamics of $z$-variables. To validate our proposition, we perform the linear stability analysis of the synchronization solution $s(t)$ and calculate the maximum transverse Lyapunov exponent $\Lambda_{max}$ as a function of coupling strength $\epsilon$. The necessary condition for the emergence of stable synchronous solution is that $\Lambda_{max}$ must be negative for varying coupling strength. Figure \ref{hr_ring_lyap} delineates that the largest transverse Lyapunov exponent $\Lambda_{max}$ crosses the abscissa at a critical coupling strength $\epsilon_{c} \approx-18.5$. Beyond $\epsilon < \epsilon_{c}$, $\Lambda_{max}$ remains always negative and results in an unbounded domain of synchronization.      
\subsubsection{Random network of HR neurons}
\begin{figure}[ht] 
	\centerline{
		\includegraphics[scale=0.12]{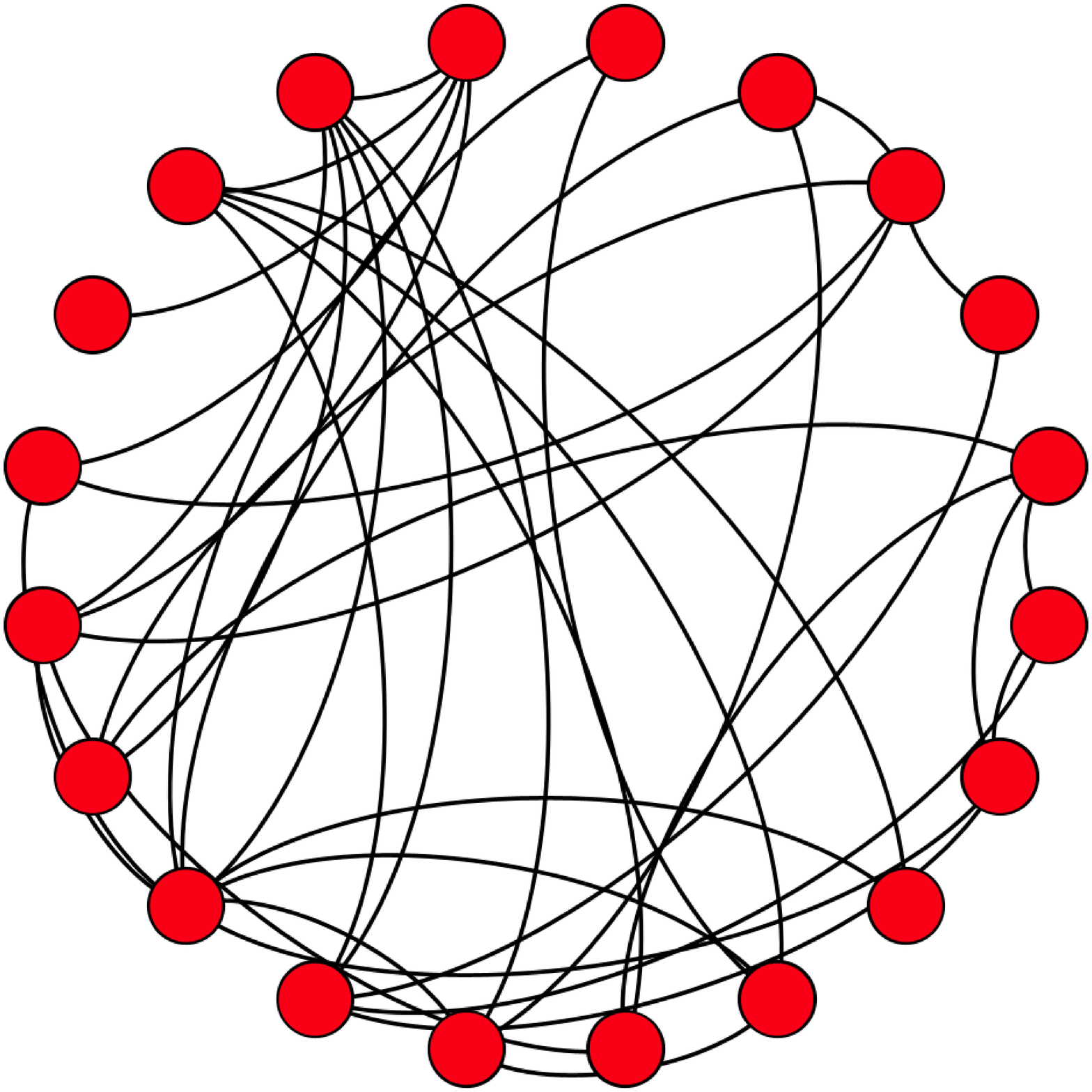}
		\includegraphics[scale=0.27]{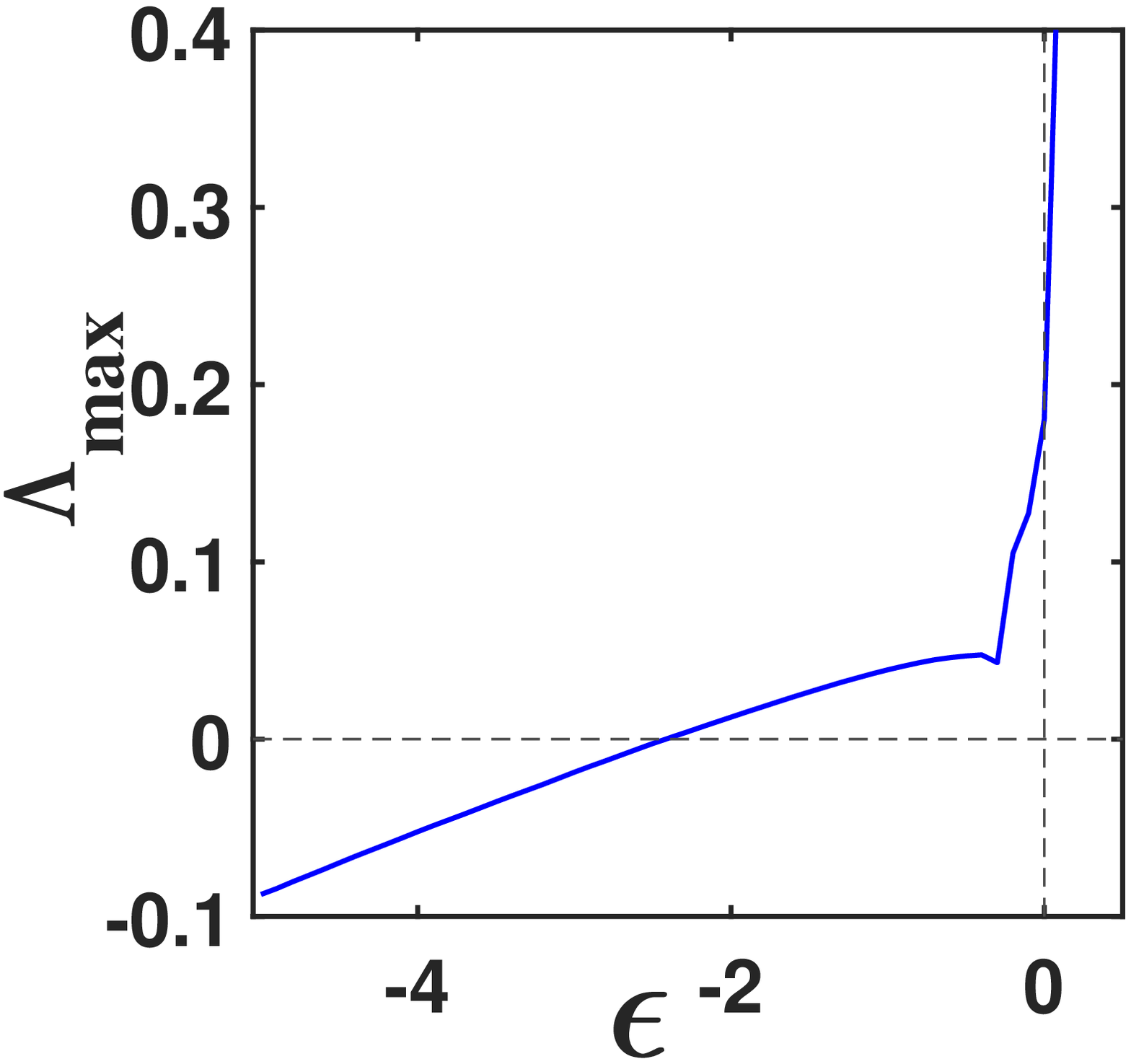}}
	\caption{ (Left panel) Schematic diagram of $N=20$ nodes random network with edge joining probability $p=0.2$. (Right panel) Variation of maximum transverse Lyapunov exponent $\Lambda_{max}$ as a function of coupling strength $\epsilon$ in the negative regime with $1 \rightarrow 3 $ coupling scheme is shown by solid blue line. For each individual HR neuron, the system parameters are $r=0.006$, $s=4$ and $I=3.2$.}
	\label{hr_random_lyap}
\end{figure} 
Lastly, we consider a complex network topology to see if our proposed coupling scheme based on the linear matrix $L$ is effective enough for the emergence of complete synchrony in the negative coupling regime with arbitrary network connectivity topology. Specifically, we consider a $N=20$ nodes Erd\H{o}s-R\'enyi random network \cite{erdos2011evolution} in which any two nodes are connected to each other with a probability $p=0.2$. The schematic representation of this random network is schematized in the left panel of Fig. \ref{hr_random_lyap}. The chaotic HR neuronal model is considered as individual node dynamics of the network and as previously based on the sign of elements of linear matrix $L$, the $1\rightarrow 3$ coupling scheme is chosen between any pair of nodes. Figure \ref{hr_random_lyap} depicts the curve of maximum transverse Lyapunov exponent $\Lambda_{max}$ for varying coupling strength $\epsilon$ in negative regime. It is observable that the curve of $\Lambda_{max}$ crosses the abscissa at a critical coupling strength $\epsilon_{c}=-2.42$ and remains negative thereafter. This reflects that our proposed coupling condition for the emergence of complete synchrony is effective even when the individuals are connected randomly with purely repulsive links.          
    \section{Conclusion} \label{conclusion}
    Summing up, we proposed a general coupling scheme that yields complete synchronization in coupled oscillators with pure repulsively coupling links. The selection of an appropriate coupling profile is done based on the nonzero off-diagonal elements of the linear matrix of the dynamical oscillators. As a result, one bidirectional cross-coupling link is introduced between any pair of oscillators, which plays the role of indicator for the achievement of complete synchrony with purely repulsive coupling. We illustrate the functioning of the proposed coupling scheme with examples of two-coupled systems using the Hindmarsh-Rose neuron model, R\"{o}ssler oscillator, Chen system, Sprott system, and NE system, as dynamical nodes. We validate the effectiveness of our result by performing linear stability analysis of the synchronous solution based on the traditional MSF scheme. We further demonstrate that our proposed coupling profiles work perfectly in network motifs of three-node, four-node, and even more, to predict the emergence of complete synchronization in these networks. An example of Erd\H{o}s-R\'enyi random network using the Hindmarsh-Rose neuron model as dynamical nodes is also exemplified that validates the effectiveness of our prescribed coupling schemes in complex networks.
    \par  Do the results obtain here for the achievement of complete synchrony in repulsively coupled dynamical oscillators carry over to all dynamical systems whose linear matrices satisfy the proposed coupling condition? There may be some instances (e.g., Sprott-H system \cite{sprott1994some}) for which the linear matrix of the system satisfies the prescribed coupling condition, but the corresponding coupled system does not show any stable complete synchronous solution. This sort of situation occurs may be because the corresponding maximum Lyapunov exponent shows type-I \cite{boccaletti2006complex} behavior, i.e., the curve of maximum Lyapunov exponent does not cross the zero line for any coupling strengths. Despite this fact, our result provides a general scheme for the achievement of complete synchrony with purely repulsive coupling in a large class of interacting dynamical systems.  We expect, our result will pave a way to understand various collective phenomena in pure repulsively coupled systems beyond canonical oscillators, for example in living systems or social systems \cite{gosak2022networks}.

\appendix
 \section{Examples of two-coupled systems} \label{appendix}
 Here, we provide few more examples of two coupled systems that show complete synchronization with purely repulsive coupling scheme. Table \ref{table:1} illustrates the corresponding results along with the appropriate coupling scheme obtained from our prescribed condition based on the linear matrices.  
\begin{longtable*}[c]{|p{4.5cm}|p{3.5cm}|p{2.2cm}|p{3.0cm}| }
	\caption{Occurrence of complete synchrony in different two-coupled systems with purely repulsive coupling.\label{table:1}}\\
	\hline \hline
	System &  linear matrix $(L)$  & Cross-coupling link & Critical coupling $(K=2\epsilon)$  \\ 
	\hline \hline
	Sprott-K : 
			\begin{eqnarray*} \dot{x}=xy-z,\\ \dot{y}=x-y, \\ \dot{z}= x+0.3z.\end{eqnarray*} & \[\begin{pmatrix} 0 & 0 & -1 \\1 & -1 & 0 \\1 & 0 & 0.3\end{pmatrix}\] & \[1\to 3\] &  \[-1.4 < K < -0.725\] \\
	\hline
	Sprott-M: 
	 \begin{eqnarray*} \dot{x}=-z, \\ \dot{y}=-x^2-y, \\ \dot{z}= 1.7+1.7x+y. \end{eqnarray*} & \[ \begin{pmatrix} 0 & 0 & -1 \\0 & -1 & 0 \\1.7 & 1 & 0\end{pmatrix} \] & \[1\to 3\] &  \[ K < -1.3625\] \\
	\hline
	Sprott-O:
	\begin{eqnarray*} \dot{x}=y, \\ \dot{y}=x-z, \\ \dot{z}=  x+xz+2.7y. \end{eqnarray*} & \[ \begin{pmatrix} 0 & 1 & 0 \\1 & 0 & -1 \\1 & 2.7 & 0\end{pmatrix} \] & \[2\to 3\] & \[ K < -0.2\]  \\
	\hline
	Sprott-P: 
	\begin{eqnarray*} \dot{x}=2.7y+z, \\ \dot{y}=-x+y^2, \\ \dot{z}= x+y. \end{eqnarray*} & \[ \begin{pmatrix} 0 & 2.7 & 1 \\-1 & 0 & 0 \\1 & 1 & 0\end{pmatrix} \] & \[2\to 1\] & \[K<-0.95\]  \\
	\hline
	Sprott-S:
	\begin{eqnarray*} \dot{x}=-x-4y, \\ \dot{y}=x+z^2, \\ \dot{z}= 1+x. \end{eqnarray*} & \[ \begin{pmatrix} -1 & -4 & 0 \\1 & 0 & 0 \\1 & 0 & 0\end{pmatrix} \] & \[1\to 2\] &  \[K<-0.5\] \\
	\hline     
\end{longtable*}
  
\bibliographystyle{apsrev4-2} 

\begin{thebibliography}{51}%
	\makeatletter
	\providecommand \@ifxundefined [1]{%
		\@ifx{#1\undefined}
	}%
	\providecommand \@ifnum [1]{%
		\ifnum #1\expandafter \@firstoftwo
		\else \expandafter \@secondoftwo
		\fi
	}%
	\providecommand \@ifx [1]{%
		\ifx #1\expandafter \@firstoftwo
		\else \expandafter \@secondoftwo
		\fi
	}%
	\providecommand \natexlab [1]{#1}%
	\providecommand \enquote  [1]{``#1''}%
	\providecommand \bibnamefont  [1]{#1}%
	\providecommand \bibfnamefont [1]{#1}%
	\providecommand \citenamefont [1]{#1}%
	\providecommand \href@noop [0]{\@secondoftwo}%
	\providecommand \href [0]{\begingroup \@sanitize@url \@href}%
	\providecommand \@href[1]{\@@startlink{#1}\@@href}%
	\providecommand \@@href[1]{\endgroup#1\@@endlink}%
	\providecommand \@sanitize@url [0]{\catcode `\\12\catcode `\$12\catcode
		`\&12\catcode `\#12\catcode `\^12\catcode `\_12\catcode `\%12\relax}%
	\providecommand \@@startlink[1]{}%
	\providecommand \@@endlink[0]{}%
	\providecommand \url  [0]{\begingroup\@sanitize@url \@url }%
	\providecommand \@url [1]{\endgroup\@href {#1}{\urlprefix }}%
	\providecommand \urlprefix  [0]{URL }%
	\providecommand \Eprint [0]{\href }%
	\providecommand \doibase [0]{https://doi.org/}%
	\providecommand \selectlanguage [0]{\@gobble}%
	\providecommand \bibinfo  [0]{\@secondoftwo}%
	\providecommand \bibfield  [0]{\@secondoftwo}%
	\providecommand \translation [1]{[#1]}%
	\providecommand \BibitemOpen [0]{}%
	\providecommand \bibitemStop [0]{}%
	\providecommand \bibitemNoStop [0]{.\EOS\space}%
	\providecommand \EOS [0]{\spacefactor3000\relax}%
	\providecommand \BibitemShut  [1]{\csname bibitem#1\endcsname}%
	\let\auto@bib@innerbib\@empty
	\bibitem [{\citenamefont {Belykh}\ \emph {et~al.}(2016)\citenamefont {Belykh},
		\citenamefont {Brister},\ and\ \citenamefont
		{Belykh}}]{belykh2016bistability}%
	\BibitemOpen
	\bibfield  {author} {\bibinfo {author} {\bibfnamefont {I.~V.}\ \bibnamefont
			{Belykh}}, \bibinfo {author} {\bibfnamefont {B.~N.}\ \bibnamefont
			{Brister}},\ and\ \bibinfo {author} {\bibfnamefont {V.~N.}\ \bibnamefont
			{Belykh}},\ }\href@noop {} {\bibfield  {journal} {\bibinfo  {journal} {Chaos:
				An Interdisciplinary Journal of Nonlinear Science}\ }\textbf {\bibinfo
			{volume} {26}},\ \bibinfo {pages} {094822} (\bibinfo {year}
		{2016})}\BibitemShut {NoStop}%
	\bibitem [{\citenamefont {D{\"o}rfler}\ \emph {et~al.}(2013)\citenamefont
		{D{\"o}rfler}, \citenamefont {Chertkov},\ and\ \citenamefont
		{Bullo}}]{dorfler2013synchronization}%
	\BibitemOpen
	\bibfield  {author} {\bibinfo {author} {\bibfnamefont {F.}~\bibnamefont
			{D{\"o}rfler}}, \bibinfo {author} {\bibfnamefont {M.}~\bibnamefont
			{Chertkov}},\ and\ \bibinfo {author} {\bibfnamefont {F.}~\bibnamefont
			{Bullo}},\ }\href@noop {} {\bibfield  {journal} {\bibinfo  {journal}
			{Proceedings of the National Academy of Sciences}\ }\textbf {\bibinfo
			{volume} {110}},\ \bibinfo {pages} {2005} (\bibinfo {year}
		{2013})}\BibitemShut {NoStop}%
	\bibitem [{\citenamefont {D{\"o}rfler}\ and\ \citenamefont
		{Bullo}(2014)}]{dorfler2014synchronization}%
	\BibitemOpen
	\bibfield  {author} {\bibinfo {author} {\bibfnamefont {F.}~\bibnamefont
			{D{\"o}rfler}}\ and\ \bibinfo {author} {\bibfnamefont {F.}~\bibnamefont
			{Bullo}},\ }\href@noop {} {\bibfield  {journal} {\bibinfo  {journal}
			{Automatica}\ }\textbf {\bibinfo {volume} {50}},\ \bibinfo {pages} {1539}
		(\bibinfo {year} {2014})}\BibitemShut {NoStop}%
	\bibitem [{\citenamefont {Olmi}(2015)}]{olmi2015chimera}%
	\BibitemOpen
	\bibfield  {author} {\bibinfo {author} {\bibfnamefont {S.}~\bibnamefont
			{Olmi}},\ }\href@noop {} {\bibfield  {journal} {\bibinfo  {journal} {Chaos:
				An Interdisciplinary Journal of Nonlinear Science}\ }\textbf {\bibinfo
			{volume} {25}},\ \bibinfo {pages} {123125} (\bibinfo {year}
		{2015})}\BibitemShut {NoStop}%
	\bibitem [{\citenamefont {Pikovsky}\ \emph {et~al.}(2001)\citenamefont
		{Pikovsky}, \citenamefont {Rosenblum},\ and\ \citenamefont
		{Kurths}}]{syn_book}%
	\BibitemOpen
	\bibfield  {author} {\bibinfo {author} {\bibfnamefont {A.}~\bibnamefont
			{Pikovsky}}, \bibinfo {author} {\bibfnamefont {M.}~\bibnamefont
			{Rosenblum}},\ and\ \bibinfo {author} {\bibfnamefont {J.}~\bibnamefont
			{Kurths}},\ }\href@noop {} {\emph {\bibinfo {title} {Synchronization: A
				Universal Concept in Nonlinear Sciences}}}\ (\bibinfo  {publisher} {Cambridge
		University Press},\ \bibinfo {address} {Cambridge},\ \bibinfo {year}
	{2001})\BibitemShut {NoStop}%
	\bibitem [{\citenamefont {Boccaletti}\ \emph {et~al.}(2002)\citenamefont
		{Boccaletti}, \citenamefont {Kurths}, \citenamefont {Osipov}, \citenamefont
		{Valladares},\ and\ \citenamefont {Zhou}}]{boccaletti2002synchronization}%
	\BibitemOpen
	\bibfield  {author} {\bibinfo {author} {\bibfnamefont {S.}~\bibnamefont
			{Boccaletti}}, \bibinfo {author} {\bibfnamefont {J.}~\bibnamefont {Kurths}},
		\bibinfo {author} {\bibfnamefont {G.}~\bibnamefont {Osipov}}, \bibinfo
		{author} {\bibfnamefont {D.}~\bibnamefont {Valladares}},\ and\ \bibinfo
		{author} {\bibfnamefont {C.}~\bibnamefont {Zhou}},\ }\href@noop {} {\bibfield
		{journal} {\bibinfo  {journal} {Physics Reports}\ }\textbf {\bibinfo
			{volume} {366}},\ \bibinfo {pages} {1} (\bibinfo {year} {2002})}\BibitemShut
	{NoStop}%
	\bibitem [{\citenamefont {Arenas}\ \emph {et~al.}(2008)\citenamefont {Arenas},
		\citenamefont {D{\'\i}az-Guilera}, \citenamefont {Kurths}, \citenamefont
		{Moreno},\ and\ \citenamefont {Zhou}}]{arenas2008synchronization}%
	\BibitemOpen
	\bibfield  {author} {\bibinfo {author} {\bibfnamefont {A.}~\bibnamefont
			{Arenas}}, \bibinfo {author} {\bibfnamefont {A.}~\bibnamefont
			{D{\'\i}az-Guilera}}, \bibinfo {author} {\bibfnamefont {J.}~\bibnamefont
			{Kurths}}, \bibinfo {author} {\bibfnamefont {Y.}~\bibnamefont {Moreno}},\
		and\ \bibinfo {author} {\bibfnamefont {C.}~\bibnamefont {Zhou}},\ }\href@noop
	{} {\bibfield  {journal} {\bibinfo  {journal} {Physics Reports}\ }\textbf
		{\bibinfo {volume} {469}},\ \bibinfo {pages} {93} (\bibinfo {year}
		{2008})}\BibitemShut {NoStop}%
	\bibitem [{\citenamefont {Pecora}\ and\ \citenamefont
		{Carroll}(1990)}]{pecora1990synchronization}%
	\BibitemOpen
	\bibfield  {author} {\bibinfo {author} {\bibfnamefont {L.~M.}\ \bibnamefont
			{Pecora}}\ and\ \bibinfo {author} {\bibfnamefont {T.~L.}\ \bibnamefont
			{Carroll}},\ }\href@noop {} {\bibfield  {journal} {\bibinfo  {journal}
			{Physical Review Letters}\ }\textbf {\bibinfo {volume} {64}},\ \bibinfo
		{pages} {821} (\bibinfo {year} {1990})}\BibitemShut {NoStop}%
	\bibitem [{\citenamefont {Cuomo}\ and\ \citenamefont
		{Oppenheim}(1993)}]{cuomo1993circuit}%
	\BibitemOpen
	\bibfield  {author} {\bibinfo {author} {\bibfnamefont {K.~M.}\ \bibnamefont
			{Cuomo}}\ and\ \bibinfo {author} {\bibfnamefont {A.~V.}\ \bibnamefont
			{Oppenheim}},\ }\href@noop {} {\bibfield  {journal} {\bibinfo  {journal}
			{Physical Review Letters}\ }\textbf {\bibinfo {volume} {71}},\ \bibinfo
		{pages} {65} (\bibinfo {year} {1993})}\BibitemShut {NoStop}%
	\bibitem [{\citenamefont {Heagy}\ \emph {et~al.}(1994)\citenamefont {Heagy},
		\citenamefont {Carroll},\ and\ \citenamefont
		{Pecora}}]{heagy1994synchronous}%
	\BibitemOpen
	\bibfield  {author} {\bibinfo {author} {\bibfnamefont {J.}~\bibnamefont
			{Heagy}}, \bibinfo {author} {\bibfnamefont {T.}~\bibnamefont {Carroll}},\
		and\ \bibinfo {author} {\bibfnamefont {L.}~\bibnamefont {Pecora}},\
	}\href@noop {} {\bibfield  {journal} {\bibinfo  {journal} {Physical Review
				E}\ }\textbf {\bibinfo {volume} {50}},\ \bibinfo {pages} {1874} (\bibinfo
		{year} {1994})}\BibitemShut {NoStop}%
	\bibitem [{\citenamefont {Rulkov}\ \emph {et~al.}(1995)\citenamefont {Rulkov},
		\citenamefont {Sushchik}, \citenamefont {Tsimring},\ and\ \citenamefont
		{Abarbanel}}]{rulkov1995generalized}%
	\BibitemOpen
	\bibfield  {author} {\bibinfo {author} {\bibfnamefont {N.~F.}\ \bibnamefont
			{Rulkov}}, \bibinfo {author} {\bibfnamefont {M.~M.}\ \bibnamefont
			{Sushchik}}, \bibinfo {author} {\bibfnamefont {L.~S.}\ \bibnamefont
			{Tsimring}},\ and\ \bibinfo {author} {\bibfnamefont {H.~D.}\ \bibnamefont
			{Abarbanel}},\ }\href@noop {} {\bibfield  {journal} {\bibinfo  {journal}
			{Physical Review E}\ }\textbf {\bibinfo {volume} {51}},\ \bibinfo {pages}
		{980} (\bibinfo {year} {1995})}\BibitemShut {NoStop}%
	\bibitem [{\citenamefont {Perc}(2009)}]{perc2009optimal}%
	\BibitemOpen
	\bibfield  {author} {\bibinfo {author} {\bibfnamefont {M.}~\bibnamefont
			{Perc}},\ }\href@noop {} {\bibfield  {journal} {\bibinfo  {journal}
			{Biophysical Chemistry}\ }\textbf {\bibinfo {volume} {141}},\ \bibinfo
		{pages} {175} (\bibinfo {year} {2009})}\BibitemShut {NoStop}%
	\bibitem [{\citenamefont {Anwar}\ and\ \citenamefont
		{Ghosh}(2022)}]{anwar2022stability}%
	\BibitemOpen
	\bibfield  {author} {\bibinfo {author} {\bibfnamefont {M.~S.}\ \bibnamefont
			{Anwar}}\ and\ \bibinfo {author} {\bibfnamefont {D.}~\bibnamefont {Ghosh}},\
	}\href@noop {} {\bibfield  {journal} {\bibinfo  {journal} {Physical Review
				E}\ }\textbf {\bibinfo {volume} {106}},\ \bibinfo {pages} {034314} (\bibinfo
		{year} {2022})}\BibitemShut {NoStop}%
	\bibitem [{\citenamefont {Mikhaylov}\ \emph {et~al.}(2013)\citenamefont
		{Mikhaylov}, \citenamefont {Komarov}, \citenamefont {Levanova},\ and\
		\citenamefont {Osipov}}]{mikhaylov2013sequential}%
	\BibitemOpen
	\bibfield  {author} {\bibinfo {author} {\bibfnamefont {A.}~\bibnamefont
			{Mikhaylov}}, \bibinfo {author} {\bibfnamefont {M.}~\bibnamefont {Komarov}},
		\bibinfo {author} {\bibfnamefont {T.}~\bibnamefont {Levanova}},\ and\
		\bibinfo {author} {\bibfnamefont {G.}~\bibnamefont {Osipov}},\ }\href@noop {}
	{\bibfield  {journal} {\bibinfo  {journal} {EPL (Europhysics Letters)}\
		}\textbf {\bibinfo {volume} {101}},\ \bibinfo {pages} {20009} (\bibinfo
		{year} {2013})}\BibitemShut {NoStop}%
	\bibitem [{\citenamefont {Belykh}\ \emph {et~al.}(2004)\citenamefont {Belykh},
		\citenamefont {Belykh},\ and\ \citenamefont {Hasler}}]{belykh2004connection}%
	\BibitemOpen
	\bibfield  {author} {\bibinfo {author} {\bibfnamefont {V.~N.}\ \bibnamefont
			{Belykh}}, \bibinfo {author} {\bibfnamefont {I.~V.}\ \bibnamefont {Belykh}},\
		and\ \bibinfo {author} {\bibfnamefont {M.}~\bibnamefont {Hasler}},\
	}\href@noop {} {\bibfield  {journal} {\bibinfo  {journal} {Physica D:
				Nonlinear Phenomena}\ }\textbf {\bibinfo {volume} {195}},\ \bibinfo {pages}
		{159} (\bibinfo {year} {2004})}\BibitemShut {NoStop}%
	\bibitem [{\citenamefont {Babaoglu}\ \emph {et~al.}(2007)\citenamefont
		{Babaoglu}, \citenamefont {Binci}, \citenamefont {Jelasity},\ and\
		\citenamefont {Montresor}}]{babaoglu2007firefly}%
	\BibitemOpen
	\bibfield  {author} {\bibinfo {author} {\bibfnamefont {O.}~\bibnamefont
			{Babaoglu}}, \bibinfo {author} {\bibfnamefont {T.}~\bibnamefont {Binci}},
		\bibinfo {author} {\bibfnamefont {M.}~\bibnamefont {Jelasity}},\ and\
		\bibinfo {author} {\bibfnamefont {A.}~\bibnamefont {Montresor}},\ }in\
	\href@noop {} {\emph {\bibinfo {booktitle} {First International Conference on
				Self-Adaptive and Self-Organizing Systems (SASO 2007)}}}\ (\bibinfo
	{organization} {IEEE},\ \bibinfo {year} {2007})\ pp.\ \bibinfo {pages}
	{77--86}\BibitemShut {NoStop}%
	\bibitem [{\citenamefont {Abraham}\ \emph {et~al.}(2017)\citenamefont
		{Abraham}, \citenamefont {Nivala} \emph {et~al.}}]{abraham2017chaotic}%
	\BibitemOpen
	\bibfield  {author} {\bibinfo {author} {\bibfnamefont {R.}~\bibnamefont
			{Abraham}}, \bibinfo {author} {\bibfnamefont {M.}~\bibnamefont {Nivala}},
		\emph {et~al.},\ }\href@noop {} {\emph {\bibinfo {title} {Chaotic
				Synchronization in Economic Networks}}},\ \bibinfo {type} {Tech. Rep.}\
	(\bibinfo  {institution} {Department of Economics, University of Siena},\
	\bibinfo {year} {2017})\BibitemShut {NoStop}%
	\bibitem [{\citenamefont {Milanovi{\'c}}\ and\ \citenamefont
		{Zaghloul}(1996)}]{milanovic1996synchronization}%
	\BibitemOpen
	\bibfield  {author} {\bibinfo {author} {\bibfnamefont {V.}~\bibnamefont
			{Milanovi{\'c}}}\ and\ \bibinfo {author} {\bibfnamefont {M.~E.}\ \bibnamefont
			{Zaghloul}},\ }\href@noop {} {\bibfield  {journal} {\bibinfo  {journal}
			{International Journal of Bifurcation and Chaos}\ }\textbf {\bibinfo {volume}
			{6}},\ \bibinfo {pages} {2571} (\bibinfo {year} {1996})}\BibitemShut
	{NoStop}%
	\bibitem [{\citenamefont {Blasius}\ and\ \citenamefont
		{Stone}(2000)}]{blasius2000chaos}%
	\BibitemOpen
	\bibfield  {author} {\bibinfo {author} {\bibfnamefont {B.}~\bibnamefont
			{Blasius}}\ and\ \bibinfo {author} {\bibfnamefont {L.}~\bibnamefont
			{Stone}},\ }\href@noop {} {\bibfield  {journal} {\bibinfo  {journal}
			{International Journal of Bifurcation and Chaos}\ }\textbf {\bibinfo {volume}
			{10}},\ \bibinfo {pages} {2361} (\bibinfo {year} {2000})}\BibitemShut
	{NoStop}%
	\bibitem [{\citenamefont {Pecora}\ and\ \citenamefont
		{Carroll}(1998)}]{pecora1998master}%
	\BibitemOpen
	\bibfield  {author} {\bibinfo {author} {\bibfnamefont {L.~M.}\ \bibnamefont
			{Pecora}}\ and\ \bibinfo {author} {\bibfnamefont {T.~L.}\ \bibnamefont
			{Carroll}},\ }\href@noop {} {\bibfield  {journal} {\bibinfo  {journal}
			{Physical Review Letters}\ }\textbf {\bibinfo {volume} {80}},\ \bibinfo
		{pages} {2109} (\bibinfo {year} {1998})}\BibitemShut {NoStop}%
	\bibitem [{\citenamefont {Anwar}\ \emph {et~al.}(2022)\citenamefont {Anwar},
		\citenamefont {Rakshit}, \citenamefont {Ghosh},\ and\ \citenamefont
		{Bollt}}]{intra2}%
	\BibitemOpen
	\bibfield  {author} {\bibinfo {author} {\bibfnamefont {M.~S.}\ \bibnamefont
			{Anwar}}, \bibinfo {author} {\bibfnamefont {S.}~\bibnamefont {Rakshit}},
		\bibinfo {author} {\bibfnamefont {D.}~\bibnamefont {Ghosh}},\ and\ \bibinfo
		{author} {\bibfnamefont {E.~M.}\ \bibnamefont {Bollt}},\ }\href@noop {}
	{\bibfield  {journal} {\bibinfo  {journal} {Physical Review E}\ }\textbf
		{\bibinfo {volume} {105}},\ \bibinfo {pages} {024303} (\bibinfo {year}
		{2022})}\BibitemShut {NoStop}%
	\bibitem [{\citenamefont {Parastesh}\ \emph {et~al.}(2019)\citenamefont
		{Parastesh}, \citenamefont {Azarnoush}, \citenamefont {Jafari}, \citenamefont
		{Hatef}, \citenamefont {Perc},\ and\ \citenamefont
		{Repnik}}]{parastesh2019synchronizability}%
	\BibitemOpen
	\bibfield  {author} {\bibinfo {author} {\bibfnamefont {F.}~\bibnamefont
			{Parastesh}}, \bibinfo {author} {\bibfnamefont {H.}~\bibnamefont
			{Azarnoush}}, \bibinfo {author} {\bibfnamefont {S.}~\bibnamefont {Jafari}},
		\bibinfo {author} {\bibfnamefont {B.}~\bibnamefont {Hatef}}, \bibinfo
		{author} {\bibfnamefont {M.}~\bibnamefont {Perc}},\ and\ \bibinfo {author}
		{\bibfnamefont {R.}~\bibnamefont {Repnik}},\ }\href@noop {} {\bibfield
		{journal} {\bibinfo  {journal} {Applied Mathematics and Computation}\
		}\textbf {\bibinfo {volume} {350}},\ \bibinfo {pages} {217} (\bibinfo {year}
		{2019})}\BibitemShut {NoStop}%
	\bibitem [{\citenamefont {Anwar}\ \emph
		{et~al.}(2021{\natexlab{a}})\citenamefont {Anwar}, \citenamefont {Kundu},\
		and\ \citenamefont {Ghosh}}]{anwar2021enhancing}%
	\BibitemOpen
	\bibfield  {author} {\bibinfo {author} {\bibfnamefont {M.~S.}\ \bibnamefont
			{Anwar}}, \bibinfo {author} {\bibfnamefont {S.}~\bibnamefont {Kundu}},\ and\
		\bibinfo {author} {\bibfnamefont {D.}~\bibnamefont {Ghosh}},\ }\href@noop {}
	{\bibfield  {journal} {\bibinfo  {journal} {Chaos, Solitons \& Fractals}\
		}\textbf {\bibinfo {volume} {142}},\ \bibinfo {pages} {110476} (\bibinfo
		{year} {2021}{\natexlab{a}})}\BibitemShut {NoStop}%
	\bibitem [{\citenamefont {Parastesh}\ \emph {et~al.}(2022)\citenamefont
		{Parastesh}, \citenamefont {Rajagopal}, \citenamefont {Jafari}, \citenamefont
		{Perc},\ and\ \citenamefont {Sch{\"o}ll}}]{parastesh2022blinking}%
	\BibitemOpen
	\bibfield  {author} {\bibinfo {author} {\bibfnamefont {F.}~\bibnamefont
			{Parastesh}}, \bibinfo {author} {\bibfnamefont {K.}~\bibnamefont
			{Rajagopal}}, \bibinfo {author} {\bibfnamefont {S.}~\bibnamefont {Jafari}},
		\bibinfo {author} {\bibfnamefont {M.}~\bibnamefont {Perc}},\ and\ \bibinfo
		{author} {\bibfnamefont {E.}~\bibnamefont {Sch{\"o}ll}},\ }\href@noop {}
	{\bibfield  {journal} {\bibinfo  {journal} {Physical Review E}\ }\textbf
		{\bibinfo {volume} {105}},\ \bibinfo {pages} {054304} (\bibinfo {year}
		{2022})}\BibitemShut {NoStop}%
	\bibitem [{\citenamefont {Hoppensteadt}\ and\ \citenamefont
		{Izhikevich}(1997)}]{hoppensteadt1997weakly}%
	\BibitemOpen
	\bibfield  {author} {\bibinfo {author} {\bibfnamefont {F.~C.}\ \bibnamefont
			{Hoppensteadt}}\ and\ \bibinfo {author} {\bibfnamefont {E.~M.}\ \bibnamefont
			{Izhikevich}},\ }\href@noop {} {\emph {\bibinfo {title} {Weakly connected
				neural networks}}},\ Vol.\ \bibinfo {volume} {126}\ (\bibinfo  {publisher}
	{Springer Science \& Business Media},\ \bibinfo {year} {1997})\BibitemShut
	{NoStop}%
	\bibitem [{\citenamefont {Kim}\ \emph {et~al.}(2004)\citenamefont {Kim},
		\citenamefont {Ko}, \citenamefont {Jeong},\ and\ \citenamefont
		{Moon}}]{kim2004pattern}%
	\BibitemOpen
	\bibfield  {author} {\bibinfo {author} {\bibfnamefont {P.-J.}\ \bibnamefont
			{Kim}}, \bibinfo {author} {\bibfnamefont {T.-W.}\ \bibnamefont {Ko}},
		\bibinfo {author} {\bibfnamefont {H.}~\bibnamefont {Jeong}},\ and\ \bibinfo
		{author} {\bibfnamefont {H.-T.}\ \bibnamefont {Moon}},\ }\href@noop {}
	{\bibfield  {journal} {\bibinfo  {journal} {Physical Review E}\ }\textbf
		{\bibinfo {volume} {70}},\ \bibinfo {pages} {065201} (\bibinfo {year}
		{2004})}\BibitemShut {NoStop}%
	\bibitem [{\citenamefont {Huang}\ \emph {et~al.}(2009)\citenamefont {Huang},
		\citenamefont {Chen}, \citenamefont {Lai},\ and\ \citenamefont
		{Pecora}}]{huang2009generic}%
	\BibitemOpen
	\bibfield  {author} {\bibinfo {author} {\bibfnamefont {L.}~\bibnamefont
			{Huang}}, \bibinfo {author} {\bibfnamefont {Q.}~\bibnamefont {Chen}},
		\bibinfo {author} {\bibfnamefont {Y.-C.}\ \bibnamefont {Lai}},\ and\ \bibinfo
		{author} {\bibfnamefont {L.~M.}\ \bibnamefont {Pecora}},\ }\href@noop {}
	{\bibfield  {journal} {\bibinfo  {journal} {Physical Review E}\ }\textbf
		{\bibinfo {volume} {80}},\ \bibinfo {pages} {036204} (\bibinfo {year}
		{2009})}\BibitemShut {NoStop}%
	\bibitem [{\citenamefont {Anwar}\ \emph
		{et~al.}(2021{\natexlab{b}})\citenamefont {Anwar}, \citenamefont {Ghosh},\
		and\ \citenamefont {Frolov}}]{anwar2021relay}%
	\BibitemOpen
	\bibfield  {author} {\bibinfo {author} {\bibfnamefont {M.~S.}\ \bibnamefont
			{Anwar}}, \bibinfo {author} {\bibfnamefont {D.}~\bibnamefont {Ghosh}},\ and\
		\bibinfo {author} {\bibfnamefont {N.}~\bibnamefont {Frolov}},\ }\href@noop {}
	{\bibfield  {journal} {\bibinfo  {journal} {Mathematics}\ }\textbf {\bibinfo
			{volume} {9}},\ \bibinfo {pages} {2135} (\bibinfo {year}
		{2021}{\natexlab{b}})}\BibitemShut {NoStop}%
	\bibitem [{\citenamefont {Majhi}\ \emph {et~al.}(2020)\citenamefont {Majhi},
		\citenamefont {Chowdhury},\ and\ \citenamefont
		{Ghosh}}]{majhi2020perspective}%
	\BibitemOpen
	\bibfield  {author} {\bibinfo {author} {\bibfnamefont {S.}~\bibnamefont
			{Majhi}}, \bibinfo {author} {\bibfnamefont {S.~N.}\ \bibnamefont
			{Chowdhury}},\ and\ \bibinfo {author} {\bibfnamefont {D.}~\bibnamefont
			{Ghosh}},\ }\href@noop {} {\bibfield  {journal} {\bibinfo  {journal}
			{Europhysics Letters}\ }\textbf {\bibinfo {volume} {132}},\ \bibinfo {pages}
		{20001} (\bibinfo {year} {2020})}\BibitemShut {NoStop}%
	\bibitem [{\citenamefont {Rabinovich}\ \emph {et~al.}(2006)\citenamefont
		{Rabinovich}, \citenamefont {Varona}, \citenamefont {Selverston},\ and\
		\citenamefont {Abarbanel}}]{rabinovich2006dynamical}%
	\BibitemOpen
	\bibfield  {author} {\bibinfo {author} {\bibfnamefont {M.~I.}\ \bibnamefont
			{Rabinovich}}, \bibinfo {author} {\bibfnamefont {P.}~\bibnamefont {Varona}},
		\bibinfo {author} {\bibfnamefont {A.~I.}\ \bibnamefont {Selverston}},\ and\
		\bibinfo {author} {\bibfnamefont {H.~D.}\ \bibnamefont {Abarbanel}},\
	}\href@noop {} {\bibfield  {journal} {\bibinfo  {journal} {Reviews of Modern
				Physics}\ }\textbf {\bibinfo {volume} {78}},\ \bibinfo {pages} {1213}
		(\bibinfo {year} {2006})}\BibitemShut {NoStop}%
	\bibitem [{\citenamefont {Saha}\ \emph {et~al.}(2017)\citenamefont {Saha},
		\citenamefont {Mishra}, \citenamefont {Padmanaban}, \citenamefont {Bhowmick},
		\citenamefont {Roy}, \citenamefont {Dam},\ and\ \citenamefont
		{Dana}}]{saha2017coupling}%
	\BibitemOpen
	\bibfield  {author} {\bibinfo {author} {\bibfnamefont {S.}~\bibnamefont
			{Saha}}, \bibinfo {author} {\bibfnamefont {A.}~\bibnamefont {Mishra}},
		\bibinfo {author} {\bibfnamefont {E.}~\bibnamefont {Padmanaban}}, \bibinfo
		{author} {\bibfnamefont {S.~K.}\ \bibnamefont {Bhowmick}}, \bibinfo {author}
		{\bibfnamefont {P.~K.}\ \bibnamefont {Roy}}, \bibinfo {author} {\bibfnamefont
			{B.}~\bibnamefont {Dam}},\ and\ \bibinfo {author} {\bibfnamefont {S.~K.}\
			\bibnamefont {Dana}},\ }\href@noop {} {\bibfield  {journal} {\bibinfo
			{journal} {Physical Review E}\ }\textbf {\bibinfo {volume} {95}},\ \bibinfo
		{pages} {062204} (\bibinfo {year} {2017})}\BibitemShut {NoStop}%
	\bibitem [{\citenamefont {Saha}(2022)}]{saha2022resilience}%
	\BibitemOpen
	\bibfield  {author} {\bibinfo {author} {\bibfnamefont {S.}~\bibnamefont
			{Saha}},\ }\href@noop {} {\bibfield  {journal} {\bibinfo  {journal} {IEEE
				Transactions on Network Science and Engineering}\ }\textbf {\bibinfo {volume}
			{9}},\ \bibinfo {pages} {1594} (\bibinfo {year} {2022})}\BibitemShut
	{NoStop}%
	\bibitem [{\citenamefont {Dixit}\ \emph {et~al.}(2019)\citenamefont {Dixit},
		\citenamefont {Sharma},\ and\ \citenamefont {Shrimali}}]{dixit2019dynamics}%
	\BibitemOpen
	\bibfield  {author} {\bibinfo {author} {\bibfnamefont {S.}~\bibnamefont
			{Dixit}}, \bibinfo {author} {\bibfnamefont {A.}~\bibnamefont {Sharma}},\ and\
		\bibinfo {author} {\bibfnamefont {M.~D.}\ \bibnamefont {Shrimali}},\
	}\href@noop {} {\bibfield  {journal} {\bibinfo  {journal} {Physics Letters
				A}\ }\textbf {\bibinfo {volume} {383}},\ \bibinfo {pages} {125930} (\bibinfo
		{year} {2019})}\BibitemShut {NoStop}%
	\bibitem [{\citenamefont {Dixit}\ and\ \citenamefont
		{Shrimali}(2020)}]{dixit2020static}%
	\BibitemOpen
	\bibfield  {author} {\bibinfo {author} {\bibfnamefont {S.}~\bibnamefont
			{Dixit}}\ and\ \bibinfo {author} {\bibfnamefont {M.~D.}\ \bibnamefont
			{Shrimali}},\ }\href@noop {} {\bibfield  {journal} {\bibinfo  {journal}
			{Chaos: An Interdisciplinary Journal of Nonlinear Science}\ }\textbf
		{\bibinfo {volume} {30}},\ \bibinfo {pages} {033114} (\bibinfo {year}
		{2020})}\BibitemShut {NoStop}%
	\bibitem [{\citenamefont {Xu}\ \emph {et~al.}(2021)\citenamefont {Xu},
		\citenamefont {Tang}, \citenamefont {L{\"u}}, \citenamefont {Alfaro-Bittner},
		\citenamefont {Boccaletti}, \citenamefont {Perc},\ and\ \citenamefont
		{Guan}}]{xu2021collective}%
	\BibitemOpen
	\bibfield  {author} {\bibinfo {author} {\bibfnamefont {C.}~\bibnamefont
			{Xu}}, \bibinfo {author} {\bibfnamefont {X.}~\bibnamefont {Tang}}, \bibinfo
		{author} {\bibfnamefont {H.}~\bibnamefont {L{\"u}}}, \bibinfo {author}
		{\bibfnamefont {K.}~\bibnamefont {Alfaro-Bittner}}, \bibinfo {author}
		{\bibfnamefont {S.}~\bibnamefont {Boccaletti}}, \bibinfo {author}
		{\bibfnamefont {M.}~\bibnamefont {Perc}},\ and\ \bibinfo {author}
		{\bibfnamefont {S.}~\bibnamefont {Guan}},\ }\href@noop {} {\bibfield
		{journal} {\bibinfo  {journal} {Physical Review Research}\ }\textbf {\bibinfo
			{volume} {3}},\ \bibinfo {pages} {043004} (\bibinfo {year}
		{2021})}\BibitemShut {NoStop}%
	\bibitem [{\citenamefont {Yang}\ \emph {et~al.}(2019)\citenamefont {Yang},
		\citenamefont {Jiang},\ and\ \citenamefont {Liu}}]{yang2019synchronization}%
	\BibitemOpen
	\bibfield  {author} {\bibinfo {author} {\bibfnamefont {L.-X.}\ \bibnamefont
			{Yang}}, \bibinfo {author} {\bibfnamefont {J.}~\bibnamefont {Jiang}},\ and\
		\bibinfo {author} {\bibfnamefont {X.-J.}\ \bibnamefont {Liu}},\ }\href@noop
	{} {\bibfield  {journal} {\bibinfo  {journal} {Physica A: Statistical
				Mechanics and its Applications}\ }\textbf {\bibinfo {volume} {514}},\
		\bibinfo {pages} {916} (\bibinfo {year} {2019})}\BibitemShut {NoStop}%
	\bibitem [{\citenamefont {Leyva}\ \emph {et~al.}(2006)\citenamefont {Leyva},
		\citenamefont {Sendina-Nadal}, \citenamefont {Almendral},\ and\ \citenamefont
		{Sanju{\'a}n}}]{leyva2006sparse}%
	\BibitemOpen
	\bibfield  {author} {\bibinfo {author} {\bibfnamefont {I.}~\bibnamefont
			{Leyva}}, \bibinfo {author} {\bibfnamefont {I.}~\bibnamefont
			{Sendina-Nadal}}, \bibinfo {author} {\bibfnamefont {J.}~\bibnamefont
			{Almendral}},\ and\ \bibinfo {author} {\bibfnamefont {M.}~\bibnamefont
			{Sanju{\'a}n}},\ }\href@noop {} {\bibfield  {journal} {\bibinfo  {journal}
			{Physical Review E}\ }\textbf {\bibinfo {volume} {74}},\ \bibinfo {pages}
		{056112} (\bibinfo {year} {2006})}\BibitemShut {NoStop}%
	\bibitem [{\citenamefont {Kovalenko}\ \emph {et~al.}(2021)\citenamefont
		{Kovalenko}, \citenamefont {Dai}, \citenamefont {Alfaro-Bittner},
		\citenamefont {Raigorodskii}, \citenamefont {Perc},\ and\ \citenamefont
		{Boccaletti}}]{kovalenko2021contrarians}%
	\BibitemOpen
	\bibfield  {author} {\bibinfo {author} {\bibfnamefont {K.}~\bibnamefont
			{Kovalenko}}, \bibinfo {author} {\bibfnamefont {X.}~\bibnamefont {Dai}},
		\bibinfo {author} {\bibfnamefont {K.}~\bibnamefont {Alfaro-Bittner}},
		\bibinfo {author} {\bibfnamefont {A.}~\bibnamefont {Raigorodskii}}, \bibinfo
		{author} {\bibfnamefont {M.}~\bibnamefont {Perc}},\ and\ \bibinfo {author}
		{\bibfnamefont {S.}~\bibnamefont {Boccaletti}},\ }\href@noop {} {\bibfield
		{journal} {\bibinfo  {journal} {Physical Review Letters}\ }\textbf {\bibinfo
			{volume} {127}},\ \bibinfo {pages} {258301} (\bibinfo {year}
		{2021})}\BibitemShut {NoStop}%
	\bibitem [{\citenamefont {Elson}\ \emph {et~al.}(1998)\citenamefont {Elson},
		\citenamefont {Selverston}, \citenamefont {Huerta}, \citenamefont {Rulkov},
		\citenamefont {Rabinovich},\ and\ \citenamefont
		{Abarbanel}}]{elson1998synchronous}%
	\BibitemOpen
	\bibfield  {author} {\bibinfo {author} {\bibfnamefont {R.~C.}\ \bibnamefont
			{Elson}}, \bibinfo {author} {\bibfnamefont {A.~I.}\ \bibnamefont
			{Selverston}}, \bibinfo {author} {\bibfnamefont {R.}~\bibnamefont {Huerta}},
		\bibinfo {author} {\bibfnamefont {N.~F.}\ \bibnamefont {Rulkov}}, \bibinfo
		{author} {\bibfnamefont {M.~I.}\ \bibnamefont {Rabinovich}},\ and\ \bibinfo
		{author} {\bibfnamefont {H.~D.}\ \bibnamefont {Abarbanel}},\ }\href@noop {}
	{\bibfield  {journal} {\bibinfo  {journal} {Physical Review Letters}\
		}\textbf {\bibinfo {volume} {81}},\ \bibinfo {pages} {5692} (\bibinfo {year}
		{1998})}\BibitemShut {NoStop}%
	\bibitem [{\citenamefont {Menck}\ \emph {et~al.}(2013)\citenamefont {Menck},
		\citenamefont {Heitzig}, \citenamefont {Marwan},\ and\ \citenamefont
		{Kurths}}]{menck2013basin}%
	\BibitemOpen
	\bibfield  {author} {\bibinfo {author} {\bibfnamefont {P.~J.}\ \bibnamefont
			{Menck}}, \bibinfo {author} {\bibfnamefont {J.}~\bibnamefont {Heitzig}},
		\bibinfo {author} {\bibfnamefont {N.}~\bibnamefont {Marwan}},\ and\ \bibinfo
		{author} {\bibfnamefont {J.}~\bibnamefont {Kurths}},\ }\href@noop {}
	{\bibfield  {journal} {\bibinfo  {journal} {Nature Physics}\ }\textbf
		{\bibinfo {volume} {9}},\ \bibinfo {pages} {89} (\bibinfo {year}
		{2013})}\BibitemShut {NoStop}%
	\bibitem [{\citenamefont {Rakshit}\ \emph {et~al.}(2017)\citenamefont
		{Rakshit}, \citenamefont {Bera}, \citenamefont {Majhi}, \citenamefont
		{Hens},\ and\ \citenamefont {Ghosh}}]{rakshit2017basin}%
	\BibitemOpen
	\bibfield  {author} {\bibinfo {author} {\bibfnamefont {S.}~\bibnamefont
			{Rakshit}}, \bibinfo {author} {\bibfnamefont {B.~K.}\ \bibnamefont {Bera}},
		\bibinfo {author} {\bibfnamefont {S.}~\bibnamefont {Majhi}}, \bibinfo
		{author} {\bibfnamefont {C.}~\bibnamefont {Hens}},\ and\ \bibinfo {author}
		{\bibfnamefont {D.}~\bibnamefont {Ghosh}},\ }\href@noop {} {\bibfield
		{journal} {\bibinfo  {journal} {Scientific Reports}\ }\textbf {\bibinfo
			{volume} {7}},\ \bibinfo {pages} {1} (\bibinfo {year} {2017})}\BibitemShut
	{NoStop}%
	\bibitem [{\citenamefont {Chen}\ and\ \citenamefont
		{Ueta}(1999)}]{chen1999yet}%
	\BibitemOpen
	\bibfield  {author} {\bibinfo {author} {\bibfnamefont {G.}~\bibnamefont
			{Chen}}\ and\ \bibinfo {author} {\bibfnamefont {T.}~\bibnamefont {Ueta}},\
	}\href@noop {} {\bibfield  {journal} {\bibinfo  {journal} {International
				Journal of Bifurcation and Chaos}\ }\textbf {\bibinfo {volume} {9}},\
		\bibinfo {pages} {1465} (\bibinfo {year} {1999})}\BibitemShut {NoStop}%
	\bibitem [{\citenamefont {Sprott}(1994)}]{sprott1994some}%
	\BibitemOpen
	\bibfield  {author} {\bibinfo {author} {\bibfnamefont {J.~C.}\ \bibnamefont
			{Sprott}},\ }\href@noop {} {\bibfield  {journal} {\bibinfo  {journal}
			{Physical Review E}\ }\textbf {\bibinfo {volume} {50}},\ \bibinfo {pages}
		{R647} (\bibinfo {year} {1994})}\BibitemShut {NoStop}%
	\bibitem [{spr()}]{sprottI}%
	\BibitemOpen
	\href@noop {} {\bibinfo  {journal} {Sprott-I system: {$\dot{x}=-0.2y$,
				$\dot{y}=x+z$, $\dot{z}=x+y^{2}-z$}}\ }\BibitemShut {NoStop}%
	\bibitem [{\citenamefont {R{\"o}ssler}(1976)}]{rossler1976equation}%
	\BibitemOpen
	\bibfield  {journal} {  }\bibfield  {author} {\bibinfo {author} {\bibfnamefont
			{O.~E.}\ \bibnamefont {R{\"o}ssler}},\ }\href@noop {} {\bibfield  {journal}
		{\bibinfo  {journal} {Physics Letters A}\ }\textbf {\bibinfo {volume} {57}},\
		\bibinfo {pages} {397} (\bibinfo {year} {1976})}\BibitemShut {NoStop}%
	\bibitem [{ros()}]{rossler}%
	\BibitemOpen
	\href@noop {} {\bibinfo  {journal} {R\"{o}ssler oscillator: {$\dot{x}=-y-z$,
				$\dot{y}=x+ay$, $\dot{z}=b+z(x-c)$}, parameters: {$a=b=0.2$}, {$c=5.7$}}\
	}\BibitemShut {NoStop}%
	\bibitem [{\citenamefont {Jafari}\ \emph {et~al.}(2013)\citenamefont {Jafari},
		\citenamefont {Sprott},\ and\ \citenamefont
		{Golpayegani}}]{jafari2013elementary}%
	\BibitemOpen
	\bibfield  {journal} {  }\bibfield  {author} {\bibinfo {author} {\bibfnamefont
			{S.}~\bibnamefont {Jafari}}, \bibinfo {author} {\bibfnamefont
			{J.}~\bibnamefont {Sprott}},\ and\ \bibinfo {author} {\bibfnamefont {S.~M.
				R.~H.}\ \bibnamefont {Golpayegani}},\ }\href@noop {} {\bibfield  {journal}
		{\bibinfo  {journal} {Physics Letters A}\ }\textbf {\bibinfo {volume}
			{377}},\ \bibinfo {pages} {699} (\bibinfo {year} {2013})}\BibitemShut
	{NoStop}%
	\bibitem [{ne3()}]{ne3}%
	\BibitemOpen
	\href@noop {} {\bibinfo  {journal} {NE3 sysytem: {$\dot{x}=y$, $\dot{y}=z$,
				$\dot{z}=-y+0.1x^{2}+1.1xz+1$}}\ }\BibitemShut {NoStop}%
	\bibitem [{\citenamefont {Erd{\"o}s}\ and\ \citenamefont
		{R{\'e}nyi}(2011)}]{erdos2011evolution}%
	\BibitemOpen
	\bibfield  {journal} {  }\bibfield  {author} {\bibinfo {author} {\bibfnamefont
			{P.}~\bibnamefont {Erd{\"o}s}}\ and\ \bibinfo {author} {\bibfnamefont
			{A.}~\bibnamefont {R{\'e}nyi}},\ }in\ \href@noop {} {\emph {\bibinfo
			{booktitle} {The Structure and Dynamics of Networks}}}\ (\bibinfo
	{publisher} {Princeton University Press},\ \bibinfo {year} {2011})\ pp.\
	\bibinfo {pages} {38--82}\BibitemShut {NoStop}%
	\bibitem [{\citenamefont {Boccaletti}\ \emph {et~al.}(2006)\citenamefont
		{Boccaletti}, \citenamefont {Latora}, \citenamefont {Moreno}, \citenamefont
		{Chavez},\ and\ \citenamefont {Hwang}}]{boccaletti2006complex}%
	\BibitemOpen
	\bibfield  {author} {\bibinfo {author} {\bibfnamefont {S.}~\bibnamefont
			{Boccaletti}}, \bibinfo {author} {\bibfnamefont {V.}~\bibnamefont {Latora}},
		\bibinfo {author} {\bibfnamefont {Y.}~\bibnamefont {Moreno}}, \bibinfo
		{author} {\bibfnamefont {M.}~\bibnamefont {Chavez}},\ and\ \bibinfo {author}
		{\bibfnamefont {D.-U.}\ \bibnamefont {Hwang}},\ }\href@noop {} {\bibfield
		{journal} {\bibinfo  {journal} {Physics Reports}\ }\textbf {\bibinfo {volume}
			{424}},\ \bibinfo {pages} {175} (\bibinfo {year} {2006})}\BibitemShut
	{NoStop}%
	\bibitem [{\citenamefont {Gosak}\ \emph {et~al.}(2022)\citenamefont {Gosak},
		\citenamefont {Milojevi{\'c}}, \citenamefont {Duh}, \citenamefont {Skok},\
		and\ \citenamefont {Perc}}]{gosak2022networks}%
	\BibitemOpen
	\bibfield  {author} {\bibinfo {author} {\bibfnamefont {M.}~\bibnamefont
			{Gosak}}, \bibinfo {author} {\bibfnamefont {M.}~\bibnamefont
			{Milojevi{\'c}}}, \bibinfo {author} {\bibfnamefont {M.}~\bibnamefont {Duh}},
		\bibinfo {author} {\bibfnamefont {K.}~\bibnamefont {Skok}},\ and\ \bibinfo
		{author} {\bibfnamefont {M.}~\bibnamefont {Perc}},\ }\href@noop {} {\bibfield
		{journal} {\bibinfo  {journal} {Physics of Life Reviews}\ } (\bibinfo {year}
		{2022})}\BibitemShut {NoStop}%
\end{thebibliography}
%

\end{document}